\documentclass[twocolumn]{aastex62}

\usepackage{amsmath}	
\usepackage{amssymb}	
\usepackage{enumitem}
\usepackage{verbatim} 
\usepackage[caption=false]{subfig}

\newcommand{\Ntargetsuniquetargets}{24,809}
\newcommand{\Ntargetsuniquetargetsone}{8,815}
\newcommand{\Ntargetsuniquetargetstwo}{8,920}
\newcommand{\Ntargetsuniquetargetsboth}{7,074}

\newcommand{\Nconfirmeduniquetargets}{1228}
\newcommand{\NconfirmeduniqueF}{85}
\newcommand{\NconfirmeduniqueG}{114}
\newcommand{\NconfirmeduniqueK}{184}
\newcommand{\NconfirmedMdwarfsTESS}{673}
\newcommand{\NconfirmedearlyMdwarfsTESS}{531}
\newcommand{\NconfirmedlateMdwarfsTESS}{142}
\newcommand{\NconfirmedearlyMdwarfsTESSbright}{189}
\newcommand{\NconfirmedlateMdwarfsTESSbright}{6}

\newcommand{\Nconfirmedprebchem}{14}

\newcommand{\NconfirmedprebchemM}{11}
\newcommand{\NconfirmedprebchemearlyM}{9}
\newcommand{\NconfirmedprebchemlateM}{2}
\newcommand{\NconfirmedprebchemearlyMbright}{0}

\newcommand{\Nconfirmedozonedeplperm}{100}

\newcommand{\NconfirmedozonedeplpermM}{15}

\newcommand{\Nconfirmedozonedeplcons}{22}

\newcommand{\Nconfirmedhyperflarestars}{99}

\newcommand{\Nconfirmedflares}{8695}
\newcommand{\Nconfirmedhyperflares}{140}

\newcommand{\amplmax}{16.1}

\newcommand{\bolenergymin}{$10^{31.0}$}
\newcommand{\bolenergymax}{$10^{36.9}$}
\newcommand{\bolenergymedian}{$10^{33.1}$}
\newcommand{\MCMEmin}{$10^{18.5}$}
\newcommand{\MCMEmax}{$10^{22.3}$}
\newcommand{\MCMEmedian}{$10^{20.0}$}

\newcommand{\TICone}{260506296}
\newcommand{\TIConeteff}{3189}

\newcommand{\TIConetessmag}{13.7}
\newcommand{\TIConebolenergymax}{$10^{34.7}$}
\newcommand{\TIConeamplmax}{16.1}

\newcommand{\TICtwo}{332487879}
\newcommand{\TICtwoteff}{5192}
\newcommand{\TICtworadius}{9.3}
\newcommand{\TICtwotessmag}{4.9}
\newcommand{\TICtwobolenergymax}{$10^{36.9}$}
\newcommand{\TICtwoamplmax}{1.01}
\newcommand{\TICtwoMCME}{$10^{22.1}$}

\newcommand{\NflarestarsKepler}{4041}
\newcommand{\NflarestarsKeplerxmatched}{4036}
\newcommand{\NflareearlyMdwarfsKepler}{110}
\newcommand{\NflarelateMdwarfsKepler}{4}

\newcommand{\Kepler}{\emph{Kepler}}
\newcommand{\TESS}{\emph{TESS}}

\begin{document}

\title[Stellar Flares from the First Tess Data Release]{Stellar Flares from the First Tess Data Release: Exploring a New Sample of M~dwarfs}
\shorttitle{Stellar Flares from the First Tess Data Release}

\correspondingauthor{Maximilian N.\ G{\"u}nther}
\email{maxgue@mit.edu}

\author[0000-0002-3164-9086]{Maximilian N.\ G{\"u}nther}
\affil{Department of Physics, and Kavli Institute for Astrophysics and Space Research, Massachusetts Institute of Technology, Cambridge, MA 02139, USA}
\affil{Juan Carlos Torres Fellow}

\author[0000-0002-4142-1800]{Zhuchang Zhan}
\affil{Department of Earth, Atmospheric, and Planetary Sciences, MIT, 77 Massachusetts Avenue, Cambridge, MA 02139, USA}

\author{Sara Seager}
\affil{Department of Physics, and Kavli Institute for Astrophysics and Space Research, Massachusetts Institute of Technology, Cambridge, MA 02139, USA}
\affil{Department of Earth, Atmospheric, and Planetary Sciences, MIT, 77 Massachusetts Avenue, Cambridge, MA 02139, USA}

\author[0000-0002-7180-081X]{Paul B. Rimmer}
\affil{Department of Earth Sciences, University of Cambridge, Downing Street, Cambridge, CB2 3EQ, UK}
\affil{Astrophysics Group, Cavendish Laboratory, J.J. Thomson Avenue, Cambridge CB3 0HE, UK}
\affil{MRC Laboratory of Molecular Biology, Francis Crick Ave, Cambridge CB2 0QH}
\affil{SCOL Senior Fellow}

\author[0000-0002-5147-9053]{Sukrit Ranjan}
\affil{Department of Earth, Atmospheric, and Planetary Sciences, MIT, 77 Massachusetts Avenue, Cambridge, MA 02139, USA}
\affil{SCOL Postdoctoral Fellow}

\author[0000-0002-3481-9052]{Keivan G. Stassun}
\affil{Vanderbilt University, Department of Physics \& Astronomy, 6301 Stevenson Center Ln., Nashville, TN 37235, USA}

\author[0000-0002-0582-1751]{Ryan J. Oelkers}
\affil{Vanderbilt University, Department of Physics \& Astronomy, 6301 Stevenson Center Ln., Nashville, TN 37235, USA}

\author[0000-0002-6939-9211]{Tansu Daylan}
\affil{Department of Physics, and Kavli Institute for Astrophysics and Space Research, Massachusetts Institute of Technology, Cambridge, MA 02139, USA}
\affil{Kavli Fellow}

\author{Elisabeth Newton}
\affil{Department of Physics, and Kavli Institute for Astrophysics and Space Research, Massachusetts Institute of Technology, Cambridge, MA 02139, USA}
\affil{Dartmouth College, Hanover, NH 03755, USA}
\affil{NSF Astronomy and Astrophysics Postdoctoral Fellow}

\author{Martti H. Kristiansen}
\affil{DTU Space, National Space Institute, Technical University of Denmark, Elektrovej 327, DK-2800 Lyngby, Denmark}
\affil{Brorfelde Observatory, Observator Gyldenkernes Vej 7, DK-4340 T\o{}ll\o{}se, Denmark}

\author{Katalin Olah}
\affil{Konkoly Observatory, CSFK, H-1121 Budapest, Konkoly Thege M. {'u}t 15-17, Hungary}

\author{Edward Gillen}
\affil{Astrophysics Group, Cavendish Laboratory, J.J. Thomson Avenue, Cambridge CB3 0HE, UK}
\affil{Winton Fellow}

\author{Saul Rappaport}
\affil{Department of Physics, and Kavli Institute for Astrophysics and Space Research, Massachusetts Institute of Technology, Cambridge, MA 02139, USA}

\author{George R.\ Ricker}
\affil{Department of Physics, and Kavli Institute for Astrophysics and Space Research, Massachusetts Institute of Technology, Cambridge, MA 02139, USA}

\author{Roland K.\ Vanderspek}
\affil{Department of Physics, and Kavli Institute for Astrophysics and Space Research, Massachusetts Institute of Technology, Cambridge, MA 02139, USA}

\author{David W.\ Latham}
\affil{Center for Astrophysics | Harvard \& Smithsonian, 60 Garden Street, Cambridge, MA 02138}

\author{Joshua N.\ Winn}
\affil{Department of Astrophysical Sciences, Princeton University, 4 Ivy Lane, Princeton, NJ 08544, USA}

\author{Jon M.\ Jenkins}
\affil{NASA Ames Research Center, Moffett Field, CA, 94035, USA}

\author{Ana Glidden}
\affil{Department of Earth and Planetary Sciences, MIT, 77 Massachusetts Avenue, Cambridge, MA 02139, USA}
\affil{Department of Physics, and Kavli Institute for Astrophysics and Space Research, Massachusetts Institute of Technology, Cambridge, MA 02139, USA}

\author{Michael Fausnaugh}
\affil{Department of Physics, and Kavli Institute for Astrophysics and Space Research, Massachusetts Institute of Technology, Cambridge, MA 02139, USA}

\author{Alan M. Levine}
\affil{Department of Physics, and Kavli Institute for Astrophysics and Space Research, Massachusetts Institute of Technology, Cambridge, MA 02139, USA}

\author{Jason A.\ Dittmann}
\affil{51 Pegasi b Postdoctoral Fellow}
\affil{Department of Earth, Atmospheric, and Planetary Sciences, MIT, 77 Massachusetts Avenue, Cambridge, MA 02139, USA}

\author{Samuel N.\ Quinn}
\affil{Center for Astrophysics | Harvard \& Smithsonian, 60 Garden Street, Cambridge, MA 02138}

\author{Akshata Krishnamurthy}
\affil{Department of Aeronautics and Astronautics, MIT, 77 Massachusetts Avenue, Cambridge, MA 02139, USA}

\author{Eric B. Ting}
\affil{NASA Ames Research Center, Moffett Field, CA, 94035, USA}

\begin{abstract}
We perform a study of stellar flares for the {\Ntargetsuniquetargets} stars observed with 2 minute cadence during the first two months of the TESS mission.
Flares may erode exoplanets' atmospheres and impact their habitability, but might also trigger the genesis of life around small stars.
TESS provides a new sample of bright dwarf stars in our galactic neighborhood, collecting data for thousands of M~dwarfs that might host habitable exoplanets.
Here, we use an automated search for flares accompanied by visual inspection. 
Then, our public \texttt{allesfitter} code robustly selects the appropriate model for potentially complex flares via Bayesian evidence.
We identify {\Nconfirmeduniquetargets} flaring stars, {\NconfirmedMdwarfsTESS} of which are M~dwarfs.
Among {\Nconfirmedflares} flares in total, the largest superflare increased the stellar brightness by a factor of {\amplmax}.
Bolometric flare energies range from {\bolenergymin} to {\bolenergymax} erg, with a median of {\bolenergymedian} erg.
Furthermore, we study the flare rate and energy as a function of stellar type and rotation period. 
We solidify past findings that fast rotating M~dwarfs are the most likely to flare, and that their flare amplitude is independent of the rotation period.
Finally, we link our results to criteria for prebiotic chemistry, atmospheric loss through coronal mass ejections, and ozone sterilization.
Four of our flaring M~dwarfs host exoplanet candidates alerted on by TESS, for which we discuss how these effects can impact life.
With upcoming TESS data releases, our flare analysis can be expanded to almost all bright small stars, aiding in defining criteria for exoplanet habitability.
\end{abstract}

\keywords{stars: flare, planetary systems, planets and satellites: general}

\section{Introduction}
The most extreme solar flare ever recorded, the `Carrington event', hit the Earth in 1859 \citep{Carrington1859, Hodgson1859}. It released a flare energy of $10^{32}$\,erg, and came accompanied by a coronal mass ejection (CME) which interacted with the Earth's magnetic field and led to destructive consequences.
Stellar flares like the Carrington event are explosive magnetic reconnection events in a star's magnetosphere, releasing bursts of isotropic radiation \citep[see e.g.][]{Benz2010, Shibata2016, Doyle2018}. Over short time scales of minutes to a few hours, they emit energy ranging form $10^{23}$\,erg (\textit{nanoflares}; e.g. \citealt{Parnell2000}) to $10^{33}-10^{38}$\,erg (\textit{superflares}; e.g. \citealt{Shibayama2013}). 
Most of the emission is in the X-ray spectrum, but in extreme cases (like the Carrington event) flares are also visible in white light. Large flares can thus be detected with optical photometric surveys such as those dedicated for exoplanet detection.
CMEs, on the other hand, are clouds of charged particles that get ejected into space with a constrained direction. 
Large flares are often accompanied by CMEs, but both events can also appear independently.

Flares and CMEs on stars hosting exoplanets can be even stronger and more frequent than those on the Sun, and can play a major role in planetary evolution and habitability. Flares may contribute to atmospheric erosion, destroy ozone layers on oxic planets, and act as stressors for surface life \citep[e.g.][]{Segura2003, Lammer2007, Scalo2007, Segura2010, Seager2013, Lingam2017superflares, Atri2017, Tilley2019, OmalleyJames2018}. In the most extreme scenario, intense flare activity could render the immediate planet surface uninhabitable, though life could survive in the ocean, in rocks, or under shallow layers of soil or dust \citep{Diaz2006, Kiang2007, Bryce2015, Tilley2019, Estrela2018}. More optimistically, flares may also power prebiotic chemistry, produce surface biosignatures, or serve as a source for otherwise scarce visible-light photosynthesis on planets orbiting M~dwarfs \citep{Bjoern2015, Airapetian2016, Mullan2018}. Indeed, recent work suggests that flares might be the only means for delivering the UV photons which have been proposed to be required to initiate life on exoplanets around M~dwarf stars \citep{Ranjan2017, Rimmer2018}. Finally, flares can alter the chemistry of planetary atmospheres, meaning their impact must be considered when interpreting atmospheric signals from a planetary atmosphere in search of biosignatures and geosignatures \citep{Grenfell2012, Venot2016, Airapetian2017}. In sum, it is critical to constrain the flare properties of exoplanet host stars to understand the evolution and habitability of their planets, and to robustly characterize their atmospheres.

M~dwarfs are of primary interest in the search for habitable exoplanets for several reasons. First, they constitute a large fraction ($\sim$70\%) of the stellar population \citep{Dole1964,  Henry1994, Reid2004, Covey2008}. Second, their small radii and low temperatures enable the detection and atmospheric characterization of habitable planets on short orbits \citep{Kaltenegger2009}.
For example, we know three exo-Earths in the `liquid-water habitable zone'\footnote{Note that definitions of the liquid-water habitable zone differ and depend on the planet mass, atmospheric composition, formation and migration history, tidal locking of the orbits, space weather, and other factors \citep[see e.g.][]{Huang1959, Kasting1993, Pierrehumbert2011, Kopparapu2013, Cullum2014, Kasting2014, Kopparapu2014, Ramirez2014, Cullum2016}.} of the red dwarf TRAPPIST-1 \citep{Gillon2017}, one around LHS 1140b \citep{Dittmann2017}, and one around Proxima Centauri \citep{Anglada2016}. Nevertheless, other factors affecting the habitability of these systems remain largely unknown. More information on the frequency and energy of the host stars' flares can help better characterize the habitability of these planets.

Seminal work on flares from selected stars includes the study by \citet{Lacy1976}, and was later expanded with surveys like the Sloan Digital Sky Survey \citep{York2000, Kowalski2009}.
The largest number of flare discoveries to date come from distant stars observed by the Kepler mission \citep{Borucki2010}. Extensive catalogs of stellar flares from Kepler were provided by, e.g., \citet{Walkowicz2011}, \citet{Hawley2014}, \citet{Davenport2016} and \citet{VanDoorsselaere2017}. The adjusted observing strategy of K2 allowed the study of bright M~dwarfs in more detail \citep{Stelzer2016}.
Ground-based wide-field surveys like the Next Generation Transit Survey \citep[NGTS;][]{Wheatley2018}, EVRYSCOPE \citep{Law2015}, and others have also contributed to flare catalogs for bright nearby objects \citep[e.g.][]{Jackman2018, Jackman2019, Howard2018}. Dedicated M~dwarf surveys like MEarth \citep{Nutzman2008} provide additional possibilities for flare studies for the smallest stars. For example, \citet{Mondrik2019} reported flares on 32 mid-to-late M dwarfs. 
Complementary, detailed spectroscopic monitoring campaigns of selected M~dwarfs give insight into their flares over multiple wavebands \citep[e.g. MUSCLES;][]{France2016}.
However, the number of bright M~dwarfs observed for high-precision and high-cadence flare studies is still limited, and the limited precision of ground-based photometric observations only allows for the detection of the most energetic flares.

The TESS mission \citep{Ricker2014}, launched in April 2018, provides the opportunity to study flares on early to late M~dwarfs. TESS is specifically designed to observe bright, small stars in the solar neighborhood. It will photometrically monitor tens of thousands of M~dwarfs, which are bright enough to allow the study of flares with a high signal-to-noise ratio.

Here, we present findings derived from the first two months of TESS data (i.e., sectors 1 and 2). Section~\ref{s:Observations} describes the TESS observations. Section~\ref{s:Methods} outlines our methodology to find flare candidates, vet against false alarms, and model flare data to estimate the flare energy. The results are presented in Section~\ref{s:Results}, and a discussion and outlook are provided in Sections~\ref{s:Discussion} and \ref{s:Conclusion}.

\section{Observations}
\label{s:Observations}

The primary goal of TESS is to search for transiting Earth-sized planets around nearby and bright stars. Its four 10 cm optical cameras simultaneously observe a total field of $24^\circ \times 96^\circ$. In its two year primary mission, TESS will measure light curves of over 200,000 pre-selected stars with a 2 minute cadence and of millions of stars with a 30 minute cadence.

This study is based on the short (2 minute) cadence data collected by TESS in sectors 1 and 2, which were made publicly available with the first data release in December 2018. These data contain {\Ntargetsuniquetargets} unique targets; {\Ntargetsuniquetargetsone} in sector 1 only, {\Ntargetsuniquetargetstwo} in sector 2 only, and {\Ntargetsuniquetargetsboth} in both sectors. 

The 2 minute cadence data were extracted using the Science Processing Operations Center (SPOC) pipeline, a descendant of the \Kepler{} mission pipeline \citep{Jenkins2002, Jenkins2010, Jenkins2016, Jenkins2017, Stumpe2014, Smith2012}. The SPOC pipeline is operated at the NASA Ames Research Center.

\section{Methods}
\label{s:Methods}

\subsection{Searching for flares}
\label{ss:Searching for flares}

To search for potential flaring stars, we analyze the {\Ntargetsuniquetargets} targets with TESS light curves from sectors 1 and 2. 
We start from the pre-search data conditioned simple aperture (PDC-SAP) lightcurves, which are detrended for instrument systematics.
Additionally, we detrend each orbit using a spline fit to remove any remaining long-term variations in the light curves.
We then remove strong periodicity from the lightcurves, which is generally caused by stellar variability or rotation. For this, we compute a Lomb-Scargle periodagram, and remove a periodic signal using a sine wave fit if two criteria are fulfilled: first, the signal must have a false alarm probability below 0.01; second, the standard deviation of the residuals must decrease. We repeat this process at maximum three times.

Next, we compute a running median using a global and a local view. For the global view, the local median flux and standard deviation, $\sigma$, is calculated using a 1024 data point bin (i.e. 1.4\,d). This is meant to catch the largest and longest flares. For the local view, we use a 128 data bin (i.e. 4.3~h) to catch smaller and shorter flares. We first run the global view iteratively and mask out all 3 $\sigma$ outliers until no more are detected. Afterwards, we run the local view with the same criterion and collect a list of all outliers.

\begin{figure*}[!htbp]
    \centering
    \includegraphics[width=\textwidth]{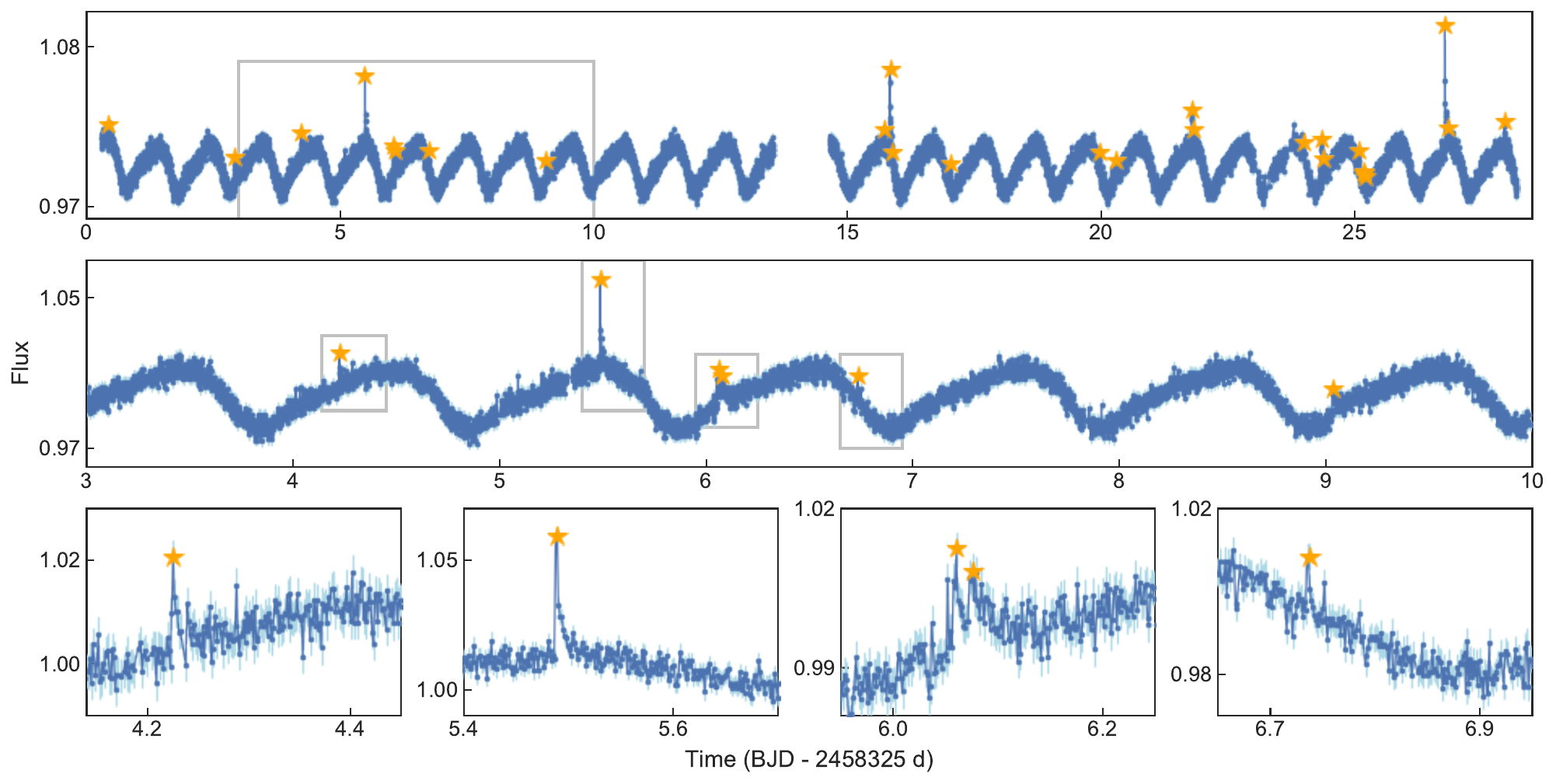}
    \caption{Demonstration of the detection pipeline on the example of the lightcurve of TIC 25118964. 
    The x-axes show the time in Barycentric Julian Days (BJD) and y-axes show the normalized TESS PDC-SAP flux. 
    Orange star symbols highlight the detected flare candidate peaks.
    The lower panels show zoomed views onto the regions marked with grey boxes.
    The flare candidate peaks are detected despite having small signal-to-noise and the presence of stellar rotation modulation. 
    }
    \label{fig:detection}
\end{figure*}

We identify outliers as flare candidates if at least six minutes of flux data lie above a 3 $\sigma$ threshold. These criteria are empirically selected to separate noise and actual flaring objects. Each target can (and often does) have multiple `flare candidate peaks'.
Fig.~\ref{fig:detection} illustrates the methodology using TIC 25118964 as an example. The flare candidate peaks are detected despite having small signal-to-noise and the presence of strong stellar rotation modulation. 
Finally we visually inspect all candidates peaks to remove false positives (e.g. asteroids) and false alarms due to noise features.

\subsection{Defining outburst epochs}
Flares are often not isolated events. When the star is active, multiple consecutive flares may occur within a short time period. We denote this collection of flares as an `outburst'. The effects of multiple flares during an outburst can overlap, resulting in compound features in the observed light curve.

We split each light curve into sections, one for each outburst epoch.
To define which flares are part of which outburst epoch, we iterate through all flare events sorted in time. 
For a given flare, if there is no other candidate peak one hour before or three hours after the event, the outburst is labeled as containing only this single flare.
When there are other flares following, the outburst epoch gets expanded accordingly.
As a result, outburst epochs span from 1~h before their earliest flare peak to 3~h after their latest flare peak.

\subsection{Completeness of the flare detection pipeline}
\label{ss:Completeness of the flare detection pipeline}

We evaluate the completeness of our flare detection pipeline using two separate injection-recovery tests. For each Sector, we randomly select lightcurves of 200 F, G and K-dwarfs, 200 early M~dwarfs and 200 late M~dwarfs. 
In the first test sample (\textit{individual flares}), we inject 10 flares into each lightcurve at random times, with amplitudes randomly drawn from a log-normal distribution between 0.01 and 1 in relative flux, and full-widths at half maximum (FWHM) randomly drawn from a uniform distribution between 2 minutes and 2 hours.
In the second test sample (\textit{outbursts}), we choose the same criteria, but force all flares to occur subsequently, mimicking extreme outburst regions. For this, we draw the first flare peak time randomly, and all subsequent peak times from a uniform distribution between 10 and 30 minutes after each other. While this is a much more extreme outburst than ever occurring in reality, it serves to test the robustness of our pipeline.

We then run our flare detection pipeline (section \ref{ss:Searching for flares}) on all injected lightcurves, and record which flares are recovered. 
In the \textit{individual flares} sample, we consider a flare recovered if a detection is registered at less than 10 minutes difference from the true peak time.
In the \textit{outbursts} sample, we consider an outburst recovered if any injected flare is recovered.

\begin{figure*}[!htbp]
    \centering
    \includegraphics[width=0.8\textwidth]{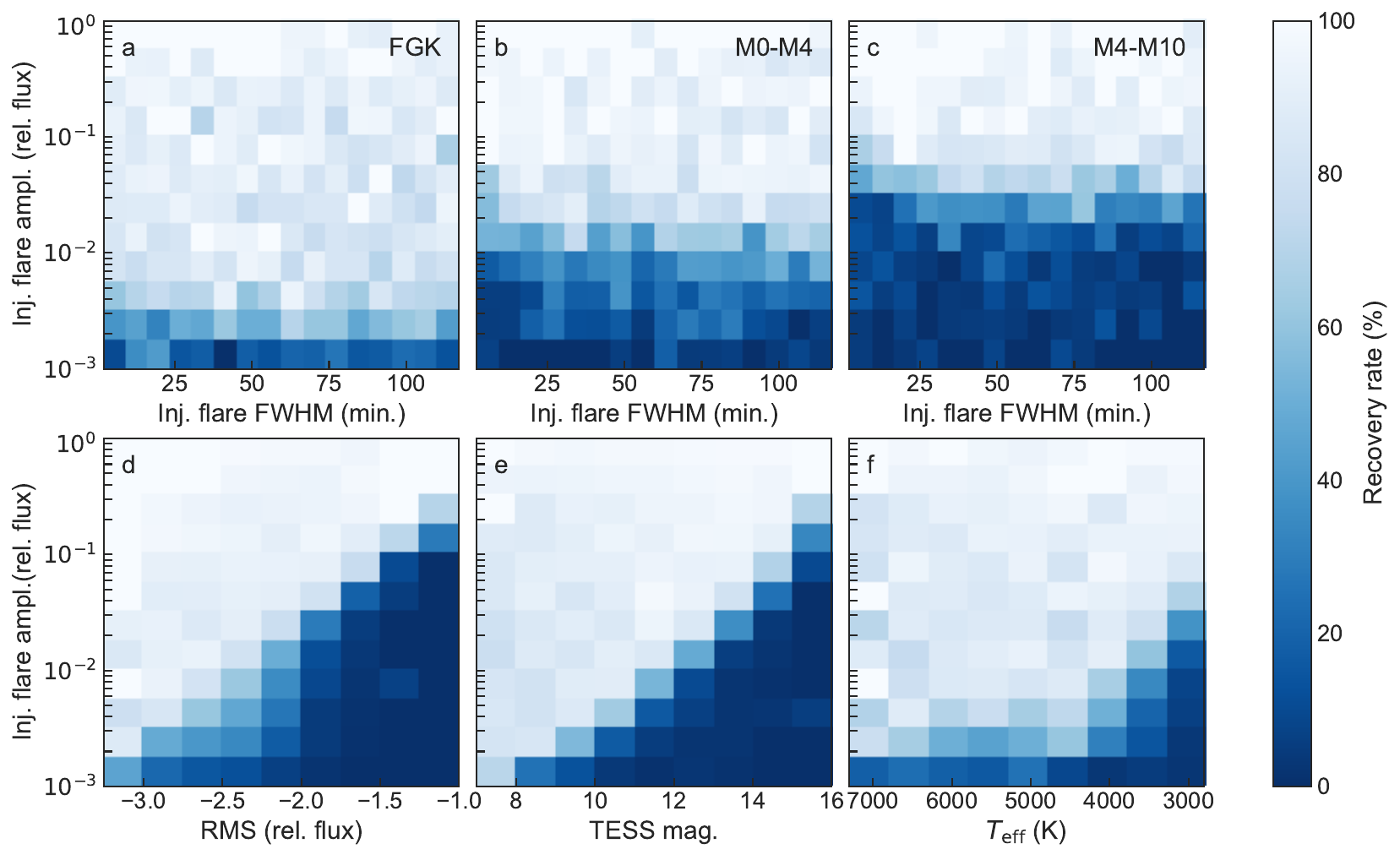}
    \caption{
    Recovery rates of flares from our injection tests. 
    Upper panels show the recovery rate as function of the injected flares' amplitudes and full-widths at half maximum (FWHM) for (a) F, G, and K-dwarfs, (b) early M~dwarfs and (c) mid/late M~dwarfs. 
    Lower panels show the recovery rate as a function of the (d) residual root-mean squared error (RMS) after detrending, (e) TESS magnitude, and (f) stellar effective temperature, all averaged over all stellar types.
    The different detection efficiency between stellar types is a direct consequence of their magnitudes, as the photometric scatter in the lightcurve increases for fainter stars.
    }
    \label{fig:plot_hist_fwhm_ampl}
\end{figure*}

We find a clear lower limit for the flare amplitude, below which flares cannot be recovered (Fig.~\ref{fig:plot_hist_fwhm_ampl}).
The recovery rate for F, G, and K-dwarfs allows the detection of the smallest injected flares, with amplitudes at the milli-mag level. 
For early M~dwarfs, we find the recovery rates decrease for flares with amplitudes of less than a few percent. 
Flares on late M~dwarfs must generally be least a few percent to be detectable. 
This dependency on stellar types is a direct consequence of the stars' brightness, as for fainter stars (such as typical late M~dwarfs) the photometric scatter (white noise) in the lightcurve increases, decreasing the signal-to-noise ratio of any flares.
The detection efficiency shows only a slight dependency on the injected flare FWHM, increasing slightly for longer flares.
Finally, as intrinsic stellar variability and rotation modulation are detrended by our flare detection pipeline, the minimum flare amplitude seems not to be affected.

\subsection{\texttt{allesfitter}}

\texttt{allesfitter} \citep{allesfitter-paper, allesfitter-code} is a publicly available, user-friendly software package for modeling photometric and RV data. It is based on a generative model that can accommodate multiple exoplanets, multi-star systems, star spots, and stellar flares.
For this, it constructs an inference framework that unites the versatile packages
\texttt{ellc} \citep[light curve and RV models;][]{Maxted2016}, 
\texttt{aflare} \citep[flare model;][]{Davenport2014},
\texttt{dynesty} \citep[static and dynamic nested sampling][]{Speagle2020},
\texttt{emcee} \citep[MCMC sampling;][]{Foreman-Mackey2013} and 
\texttt{celerite} \citep[GP models;][]{Foreman-Mackey2017}.
\texttt{allesfitter} is publicly available at \url{github.com/MNGuenther/allesfitter}, feedback and contributions are welcome.

\subsection{Modeling flares}
\label{ss:Modeling flares}

Using \texttt{allesfitter}, we perform nested sampling to infer flare models with up to $0,1, ..., N+2$ flares for each outburst epoch, where $N$ is the number of flare candidates. 
We sequentially add extra flares to ensure overlapping flares are distinguished even if they were missed by the detection pipeline and visual inspection.
We start with fitting models of `only noise' and one flare. If the model with one flare passes, we test a model with two flares, and so on, up to a maximum of $N+2$.

The model selection is performed using two complementary criteria, which both must be fulfilled. 
First, adding an extra flare must increase the logarithm of the Bayesian evidence, $\log{Z}$, by at least 5. Given a null model $M_0$, the alternative (more complex) model $M_1$ is only selected if there is sufficient relative Bayesian evidence for it as quantified by \citet{Kass1995}. Hence, we define 
\begin{align}
\Delta \log{Z} &:= \log{Z_{M_1}} - \log{Z_{M_0}},\\
\sigma (\Delta \log{Z}) &:= \sqrt{ [\sigma (\log{Z_{M_1}})]^2 - [\sigma (\log{Z_{M_0}})]^2 }
\end{align}
and demand 
\begin{align}
\Delta \log{Z} &> 5, \\
\Delta \log{Z} &> \sigma (\Delta \log{Z}).
\end{align}

Second, the extra flare must have a signal-to-noise ratio (SNR) larger than 5. For lightcurves with significant red noise, estimating the Bayesian evidence can be dominated by the volume of the prior, and be biased if flares are misused to account for noise structures (i.e. the nested sampler could find numerous solutions for placing small flares). This risk can be mitigated by introducing an SNR criterion.

We apply all fits to the detrended PDC-SAP flux. To additionally detrend any systematic noise or stellar variability features without affecting flares, we fit a Gaussian Process with a Matern 3/2 kernel in parallel with all flare models. For each model, we first run two short Markov Chain Monte Carlo (MCMC) chains to explore the likelihood space and constrain the search space for the nested sampler. This also allows us to label the sequence of flares, mitigating the risk of label-swapping during the nested sampling run.

\begin{figure*}[htbp!]
    \centering
    \subfloat[TIC~139804406]{%
      \includegraphics[width=\columnwidth]{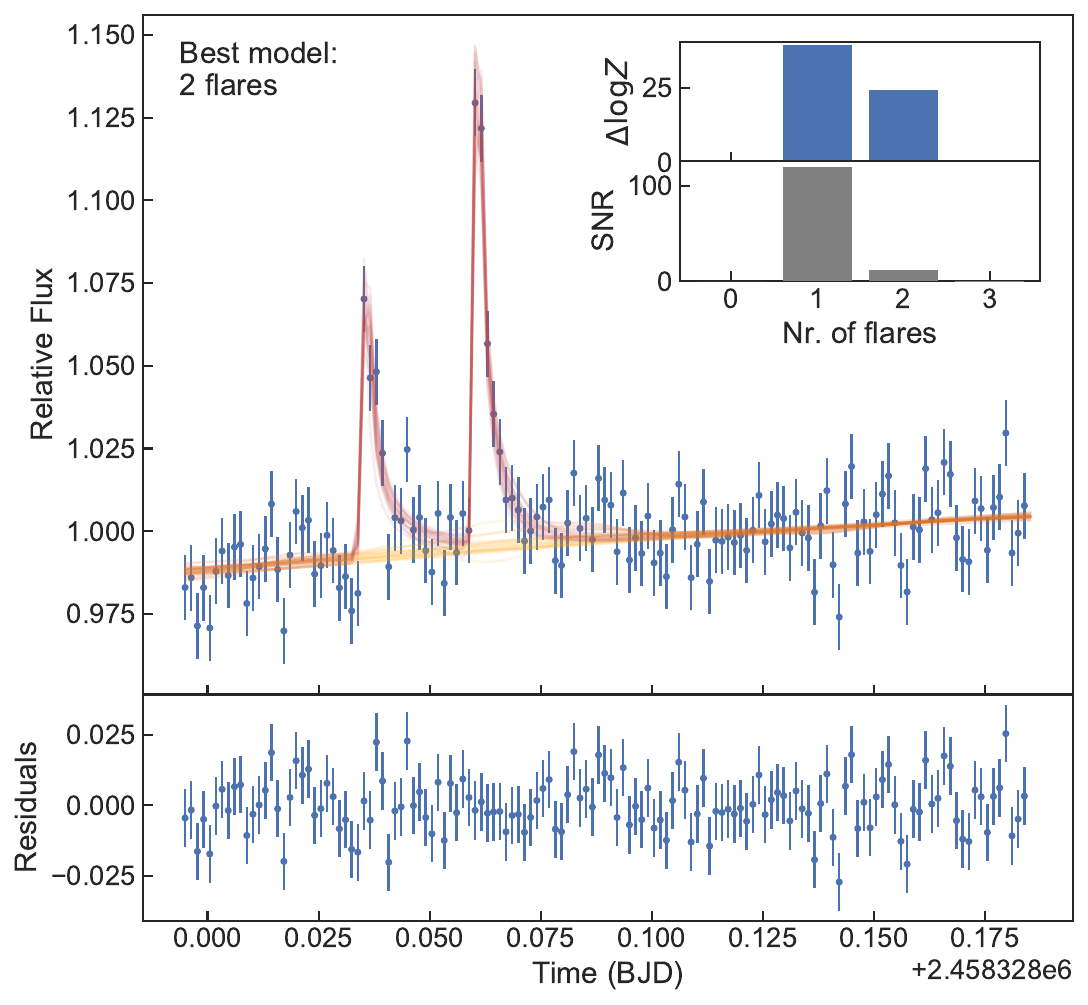}%
    }\qquad
    \subfloat[TIC~129646813]{%
      \includegraphics[width=\columnwidth]{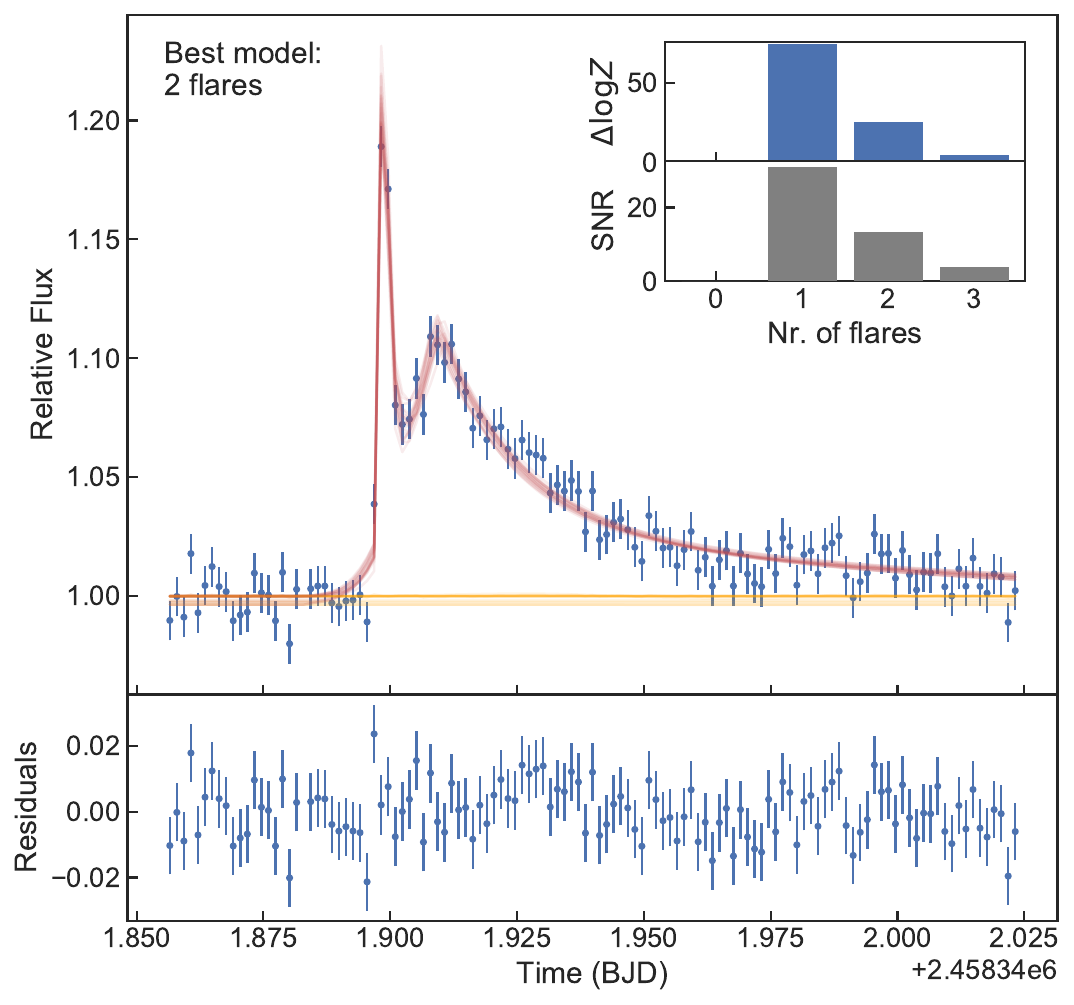}%
    }\qquad
    \subfloat[TIC~144217628]{%
      \includegraphics[width=\columnwidth]{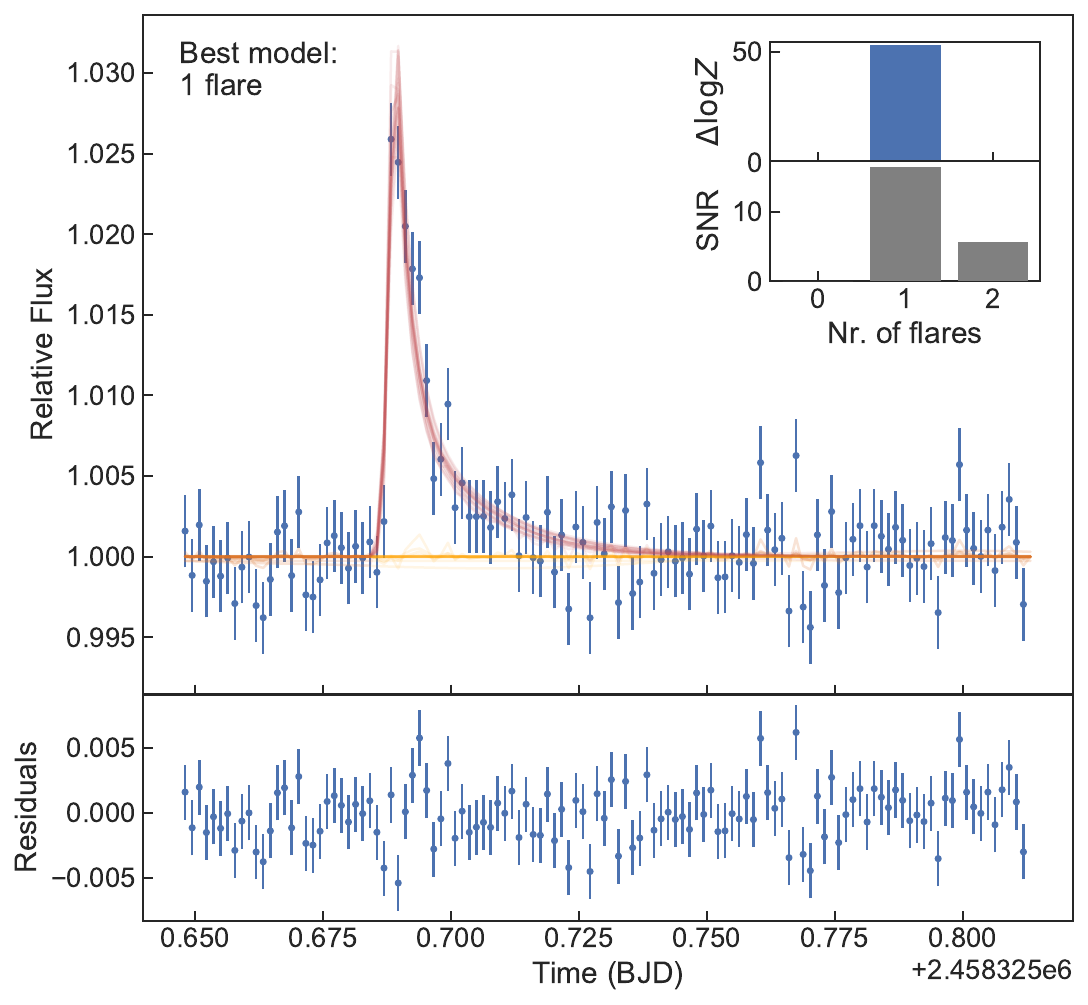}%
    }\qquad
    \subfloat[TIC~152875048]{%
      \includegraphics[width=\columnwidth]{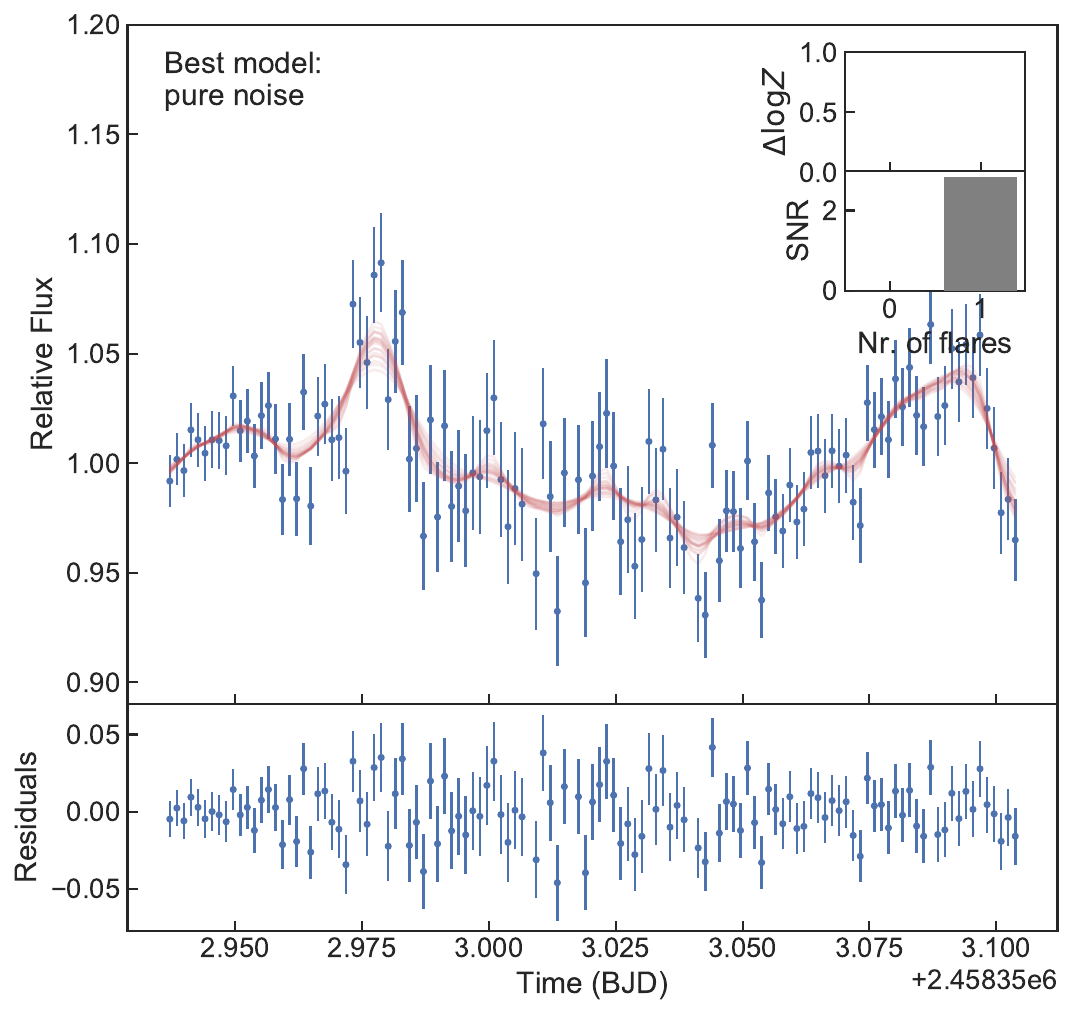}%
    }
    \caption{
    Model fits of candidate outbursts of four example stars.
    In the main figures, the x-axis shows the time in Barycentric Julian Date (BJD) and the y-axis shows the TESS PDC-SAP flux.
    Red curves show 20 posterior samples generated by the best model, while orange curves show their corresponding baselines drawn from a Gaussian Process with a Matern 3/2 kernel. 
    The curves for the less suitable models are not shown.
    The inlets show the gain in Bayesian evidence, $\Delta \log{Z}$, by adding an additional flare (upper inlay), and the SNR of the additional flare (see Section~\ref{ss:Modeling flares}).
    Initially, $N$ flare candidates were detected.
    Consequently, scenarios from $0$ to at most $N+2$ flares are fitted as long as $\Delta\log{Z}>5$ and SNR$>5$ for an additional flare.
    This allows for a quantitative model comparison, confirming the suitable number of flares (see Section~\ref{ss:Modeling flares}).
    (a) TIC~139804406: two candidate peaks were initially detected, and then confirmed using our model fit; introducing a third flare does not lead to any gain in Bayesian evidence.
    (b) TIC~129646813: only one candidate peak was detected, but our model comparison confirmed two flares; adding a third flare is not favored.
    (c) TIC~144217628: the Bayesian evidence rejects the scenario where the smaller peak is a flare; instead, it favours the peak being within the limits of a noise feature.
    (d) TIC~152875048: one candidate peak was detected, but the fit favors the pure noise model and rejects a flare scenario.
    }
    \label{fig:fit}
\end{figure*}

Fig.~\ref{fig:fit} illustrates our methodology on four example targets.
For TIC~139804406, we detected two flare candidates in the shown outburst epoch. Comparing all plausible models, we can confirm this scenario.
For TIC~129646813, the candidate list initially consisted of a single flare. However, our model fit and comparison show that there are, in fact, two flares.
For TIC~144217628, there is no evidence that the short feature towards the end of the outburst is a flare; it is best explained as a noise feature.
Finally, for TIC~52875048, the candidate peak was identified as a noise feature.

Note that the Bayesian evidence helps determine the degree to which a model is supported by the data and does \textbf{not} simply indicate its likelihood. For example, for TIC~129646813 (Fig.~\ref{fig:fit}B) we have $\Delta \log{Z_\mathrm{1 \rightarrow 2\,flares}} = 25$. This is \textbf{not} to be confused with a statement such as ``two flares are 25 times more likely than one flare". Instead, it implies that there is very strong evidence for the model with two flares.

\subsection{Vetting for asteroids and blends}
\label{ss:Vetting for asteroids and blends}

We inspected a sample of around 100 strongly flaring stars for misclassified flares caused by Solar System Objects (SSO). This effect was discussed by \cite{Pal2018} for the TESS mission. Additional catalog inputs within the same sector as the high amplitude peaks were included in our sample, making it a total of circa 600 flare events. We used \texttt{SkyBot} \citep{Berthier2006, Berthier2016} to search for known SSOs crossing the target stars at the time of flare peaks. In only a few cases we found SSOs within a range of 100''. In these instances, we carried out additional searches by slightly changing the input-time, in order to see whether the same objects would be re-detected. This was not the case in any of the computations and we therefore discards these SSOs as potential contaminants. In a single occurrence, associated with TIC 160074646 at 2458379.648 BJD$_\mathrm{TDB}$, we notice a nearby presence (21.6'') of the outer main belt SSO 2010 KM11 (519036). However, the light curve profile of this event match that of a flare profile with a fast rise and an exponential decay.
Additionally, we created a series of pixel-level animations which we then visually inspected similar to \citet{Szabo2015}, to prevent the appearance of currently unknown SSOs in the sample. No SSO-crossings were detected, including 2010 KM11 most likely due to it visual magnitude of 20.7.

In the process, we also checked for contamination by blended flaring stars. A low number of potential duplicate stars is marked in the column \textit{duplicates} in Tables~\ref{tab:catalog_per_flare} and \ref{tab:catalog_per_star}. For these, it was not possible to determine the true origin of the flaring, as the stars fall within the same pixel and potentially are part of a binary system.
We also noted a single instance for TIC~229147922 at 2458363.726719 BJD$_\mathrm{TDB}$, in which one of the flares originated from a neighboring star.

\subsection{Measuring the flare energy}
\label{ss:Measuring the flare rate}

We calculate the flare energy from the stellar luminosity and the best fitting flare profile, following \citet{Shibayama2013}.
The quiescent stellar luminosity $L_\star$ is retrieved from the TESS Input Catalog (TIC) version 8 if available.
If the TIC lists only the effective temperature $T_\mathrm{eff}$ and stellar radius $R_\star$, we compute $L_\star$ from these parameters. 
If only the $T_\mathrm{eff}$ is available, we estimate $R_\star$ using the updated values from \citet{Pecaut2013}\footnote{\url{http://www.pas.rochester.edu/~emamajek/EEM_dwarf_UBVIJHK_colors_Teff.txt}, online 28 Dec 2018} to then calculate $L_\star$.

We model the flare luminosity as black body radiation with an effective temperature of $9000\pm500$\,K, as a conservative lower limit consistent with previous studies \citep[e.g.][]{Hawley1991, Kowalski2013, Shibayama2013, Davenport2016, Jackman2018, Jackman2019, Howard2018, Chang2018}.
As discussed e.g. by \citet{Shibayama2013}, the luminosities of the star and the flare in the observing bandpass ($L'_\star$ and $L'_\mathrm{flare}$, respectively) are given as
\begin{align}
\label{eq:flare_energy_start}
L'_\star &= \pi R_\star^2 \int R_\lambda B_\lambda (T_\mathrm{eff}) \mathrm{d} \lambda \\
L'_\mathrm{flare}(t) &= A_\mathrm{flare}(t) \int R_\lambda B_\lambda (T_\mathrm{flare}) \mathrm{d} \lambda.
\end{align}
Here, $R_\lambda$ is the TESS response function, which is the product of the filter transmission and the detector quantum efficiency\footnote{\url{heasarc.gsfc.nasa.gov/docs/tess/data/tess-response-function-v1.0.csv}, online 28 Dec 2018}.
Note that the normalization of $R_\lambda$ is irrelevant here, as it cancels out in the calculations below.
$B_\lambda (T_\mathrm{eff})$ and $B_\lambda (T_\mathrm{flare})$ are the Planck functions evaluated for the star's effective temperature and the flare temperature. Last, $A_\mathrm{flare}(t)$ is the area of the flare. 
Note that this assumes that the flare temperature is constant throughout.

Since the normalized lightcurve gives us the relative flare amplitude $(\Delta F / F)(t) = L'_\mathrm{flare}(t) / L'_\star$, we can solve these equations for $A_\mathrm{flare}$:
\begin{align}
A_\mathrm{flare}(t) &= (\Delta F / F)(t) \pi R_\star^2 \frac{\int R_\lambda B_\lambda (T_\mathrm{eff}) \mathrm{d} \lambda}{\int R_\lambda B_\lambda (T_\mathrm{flare}) \mathrm{d} \lambda}.
\end{align}
This leads to the bolometric flare luminosity $L_\mathrm{flare}$:
\begin{align}
L_\mathrm{flare} &= \sigma_{\mathrm{SB}} T_\mathrm{flare}^4 A_\mathrm{flare}.
\end{align}

Finally, we arrive at the expression for the bolometric energy of the flare, given as:
\begin{align}
E_\mathrm{flare} &= \int L_\mathrm{flare}(t) \mathrm{d} t.
\label{eq:flare_energy_end}
\end{align}
We estimate the error on the flare energy by propagating all uncertainties on the stellar properties from the TESS input catalog and the estimated uncertainty on the flare temperature.

\subsection{Identifying stellar rotation periods}
\label{ss:Identifying stellar variability and rotation periods}
The rate and energy of stellar flares depend on the surface magnetic activity of the star, thus flares are thought to result from strong dynamo activity \citep[e.g.][]{Parker1979}.
Accordingly, the flaring activity should depend on the effective temperature and rotation periods (elaborated in Section~\ref{ss:Flares on fast rotators}).

We derive the stellar rotation period of all flaring and non-flaring stars from the TESS lightcurves themselves, and from a study conducted with the Kilodegree Extremely Little Telescope (KELT) by \citet{Oelkers2018}.
For TESS data, we measure the rotation period ($P_\mathrm{rot}$) using Fast Fourier Transform (FFT) computations (see \citealt{Zhan2019} for details). 
We choose the FFT method due to its robustness and computational speed. With the data being uniformly sampled, there was no advantage in using Lomb-Scargle transfoms \citep{Lomb:1976,Scargle:1982} or other methods.
We detect and analyze the FFT frequency peaks to find the primary rotation period. This is sensitive to the Nyquist limit at 4 minutes and a conservative upper limit of 5 days (to minimize false alarms due to momentum dumps and the 13-day space craft orbits). 
The reported periods and uncertainties are then determined using a Stellingwerf transform \citep{Stellingwerf1978}.
Finally, the raw light curve is phase folded onto the rotation period and examined by eye, ensuring that the signals are not caused by instrumental systematics or astrophysical false positives.

The 28~day baseline per TESS sector does not favor the identification of longer rotation periods, which are common among M dwarf stars. Studies measuring the rotation periods of nearby low-mass stars \citep{Newton2016a,Newton2018} have found a population of fast rotators ($P_\mathrm{rot}<10$\,d) and a population of slow rotators ($P_\mathrm{rot}>70$\,d), and a dearth of objects in between. Considering only their highest quality lightcurves, \citet{Newton2018} found that two-thirds of mid-to-late M dwarfs have rotation periods longer than 28 days, the length of one TESS sector, and half have periods longer than 90 days.

KELT has performed high-cadence (10--30~min), time-series photometric observations for more than four million sources since 2007. KELT observations have surveyed more than 70\% of the celestial sphere, reaching a limiting magnitude of about $V=13$, and with a baseline of nine years using KELT North and five years using KELT South. \citet{Oelkers2018} provide a catalog of $52,741$ objects showing significant photometric fluctuations likely caused by stellar variability, as determined via the Welch-Stetson $J$ and $L$ statistics \citep{Stetson:1996}. Additionally, this catalog includes $62,229$ objects identified with likely stellar rotation periods as determined by a Lomb-Scargle periodogram analysis \citep{Lomb:1976,Scargle:1982}. The detected variability ranges in root mean square (RMS) amplitude from 3~mmag to 2.3~mag, and the detected periods range from 0.1~days to over 2000~days. \citet{Oelkers2018} also provides variability upper limits for all other four million sources observed by KELT. These upper limits typically have $1\sigma$ sensitivity on 30~min timescales down to 5~mmag at $V=8$, and 43~mmag at $V=13$.

\section{Results}
\label{s:Results}

\subsection{TESS catalog of stellar flares}
\label{s:TESS catalog of stellar flares}

All detected, vetted and modeled flare events are summarized in Table~\ref{tab:catalog_per_flare} per flare event, and in Table~\ref{tab:catalog_per_star} per star.
In Table~\ref{tab:catalog_per_flare}, each star can be listed multiple times, with one row per flare. 
For each flare event, the columns show the following:
the \textit{TIC ID}; 
the TESS \textit{sector};
the \textit{outburst} number;
the \textit{flare} number;
the posterior medians for the peak time (\textit{$t_\mathrm{peak}$}), amplitude (\textit{Amp.}) and full-width at half-maximum (\textit{FWHM});
the bolometric energy of the flare (\textit{$E_\mathrm{bol.}$});
and the possible mass of a coronal mass ejection following the flare (\textit{$M_\mathrm{CME}$}; see Section~\ref{ss:The impact of coronal mass ejections}).
The machine-readable version is available in the online journal, and additionally contains lower and upper limits. For ease-of-use, it also includes a copy of the per-star columns shown in Table~\ref{tab:catalog_per_star} (see below).

In Table~\ref{tab:catalog_per_star}, each star only has one row; columns show the maximum/mean values of the flaring and the stellar parameters as follows:
the \textit{TIC ID}; 
the number of TESS sectors ($N_\mathrm{sec.}$);
the number of identified outbursts ($N_\mathrm{out.}$);
the number of identified flares ($N_\mathrm{fla.}$);
the maximum and mean flare amplitude (\textit{Amp. max.} and \textit{Amp. mean});
the maximum and mean flare full-width at half-maximum (\textit{FWHM max.} and \textit{FWHM mean});
the maximum bolometric energy of the star's flares ($E_\mathrm{bol.}$ max. and $E_\mathrm{bol.}$ mean);
the maximum and mean mass of coronal mass ejections (\textit{$M_\mathrm{CME}$} max. and \textit{$M_\mathrm{CME}$} mean);
the TESS magnitude (\textit{TESS mag});
the stellar effective temperature ($T_\mathrm{eff}$);
the stellar radius ($R_\star$);
the logarithm of the surface gravity ($\log{g}$);
the rotation period extracted from TESS and KELT data (\textit{TESS rot.} and \textit{KELT rot.}; see Section~\ref{ss:Identifying stellar variability and rotation periods}).
an indication of which star flares enough to potentially trigger prebiotic chemistry on its exoplanets (\textit{preb. chem.}; see Section~\ref{ss:Which flares deliver enough energy for prebiotic chemistry?});
and an indication of whether a star's flaring could lead to ozone depletion on its exoplanets for a conservative (\textit{ozone depl. cons.}) or permissive (\textit{ozone depl. perm.}) threshold (see Section~\ref{ss:Which flares deliver enough energy for prebiotic chemistry?});
the TIC IDs of any potential blends (\textit{duplicates}).
For readability, several columns are hidden in the printed version of Table~\ref{tab:catalog_per_star}.
The machine-readable version is available in the online journal, and contains all described columns, among all lower and upper limits values.

\subsection{TESS explores a large sample of flares on bright early and late M~dwarfs}
\label{ss:TESS explores a new flare sample: bright M dwarfs}

\begin{figure}[!htbp]
    \centering
    \includegraphics[width=\columnwidth]{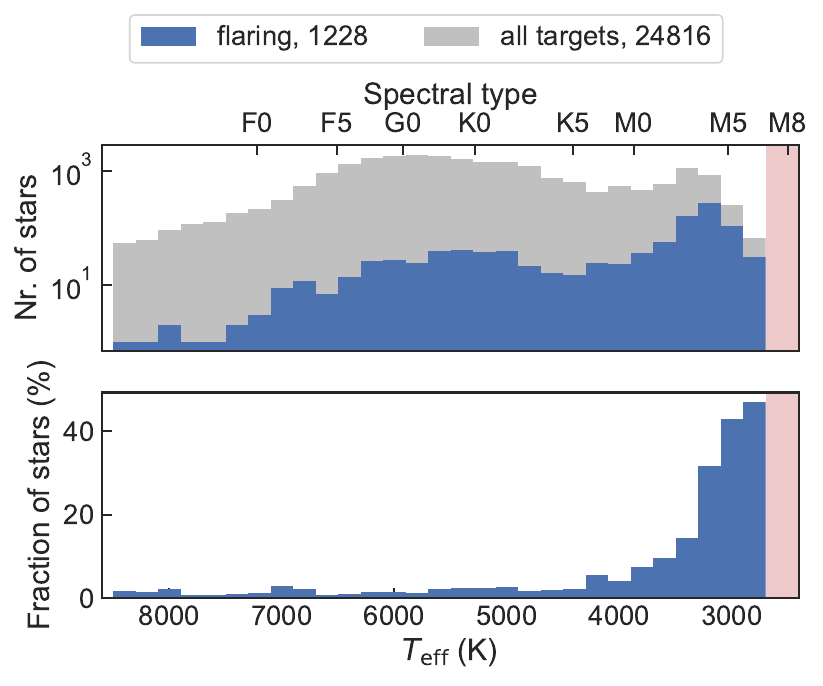}
    \caption{
    Histograms of the number (upper panel) and fraction (lower panel) of flaring stars (blue) compared with the total number of stars (grey) in the TESS short-cadence observations of sector 1 and 2; shown as a function of the stellar effective temperature $T_\mathrm{eff}$. 
    The top axis indicates stellar types following the classification by \citet{Pecaut2013}. 
    M~dwarfs dominate the sample of flaring stars, while F, G and K stars rarely have detectable flares.
    We note that for later M~dwarfs, the sample size is smaller (or zero) due to the TESS target selection.
    Additionally, the flare detection is limited by signal-to-noise constraints (see section~\ref{ss:Completeness of the flare detection pipeline}).
    }
    \label{fig:teff_vs_N}
\end{figure}

\begin{figure}[!htbp]
    \centering
    \includegraphics[width=\columnwidth]{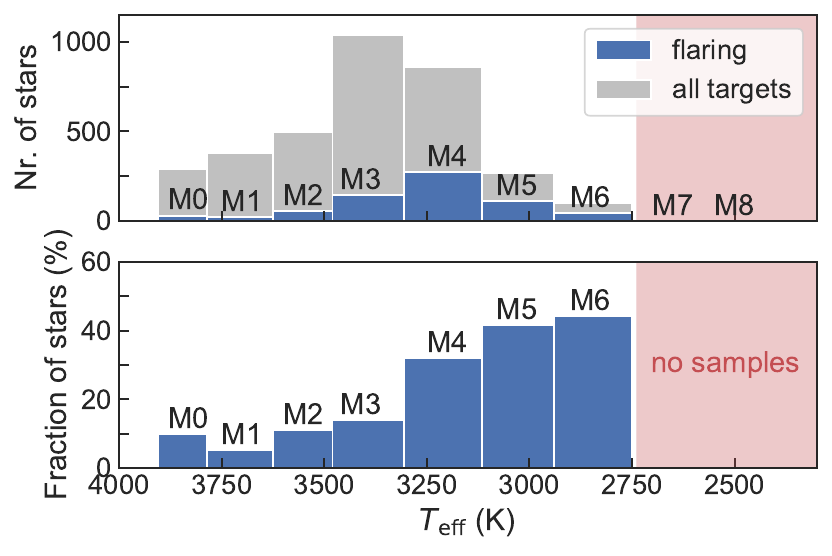}
    \caption{
    Variation of Figure~\ref{fig:teff_vs_N} focused on M~dwarfs, with bins matching the stellar types following the classification by \citet{Pecaut2013}. 
    Mid M~dwarfs of type M4--M6 constitute both the highest number (upper panel) and highest fraction (lower panel) of flaring stars, with up to 30\% of these having flares. 
    Earlier M~dwarfs seem to flare less, showing a significantly lower fraction of 5--10\%.
    Later M~dwarfs were not observed in a high enough sample size.
    We note that especially for M~dwarfs later than M4 the sample size is lower due to the TESS target selection (favoring bright stars), and the flare detection is limited by signal-to-noise constraints (see section~\ref{ss:Completeness of the flare detection pipeline}).
    }
    \label{fig:hist_of_Mdwarfs}
\end{figure}

We identify {\Nconfirmeduniquetargets} flaring stars among a total of {\Ntargetsuniquetargets} targets with short-cadence TESS observations in sectors 1 and 2 (Fig.~\ref{fig:teff_vs_N}).
This includes {\NconfirmedearlyMdwarfsTESS} flaring early M~dwarfs (M0--M4; 3905K-3200K) and {\NconfirmedlateMdwarfsTESS} flaring late M~dwarfs (M4--M10; 3200-2285K).
Fig.~\ref{fig:hist_of_Mdwarfs} shows the subsample of M~dwarfs categorized by stellar type. Flares are most commonly detected on mid M~dwarfs, which is partly influenced by the TESS target selection and signal-to-noise constraints for the flare detection. M~dwarfs of type M4--M6 show the highest fraction of flaring stars.

Mid to late M~dwarfs are the most common flare stars, with more than $40\%$ of these showing observable flares (Fig.~\ref{fig:teff_vs_N}B).
This occurrence rate is significantly lower for early M~dwarfs, as only $\sim10\%$ have observable flares.
Note that this might directly relate to the convection limit, where full convection of M dwarfs is supposed to start around spectral type M4 \citep[e.g.][]{Stassun2011}.
Hotter stars of type K and G seem to rarely host flaring events large enough to be detected (only $\sim5\%$ of these stars in our sample). 
We find flares only on {\NconfirmeduniqueF} F, {\NconfirmeduniqueG} G, and {\NconfirmeduniqueK} K stars, all of which have low amplitudes.
This confirms previous observations that M~dwarfs flare more frequently and stronger than F, G and K stars.

However, we note the interplay of two detection biases. 
Flares on M~dwarfs have a strong white light contribution and a stronger contrast against their red spectra and low quiescent luminosity. 
This enhances M~dwarfs observable amplitudes in the TESS band, favoring the detection of less energetic flares.
On the other hand, M~dwarfs also have higher photometric noise, which in turn decreases the SNR of any flares (see Section~\ref{ss:Completeness of the flare detection pipeline}).
For F, G, and K dwarfs, on the other hand, only the most energetic flares cause high enough amplitudes to be detected in contrast to their quiescent brightness in white light.
Yet, favorable for these stars is their lower photometric noise in the TESS band.

\begin{figure}[!htbp]
    \centering
    \includegraphics[width=\columnwidth]{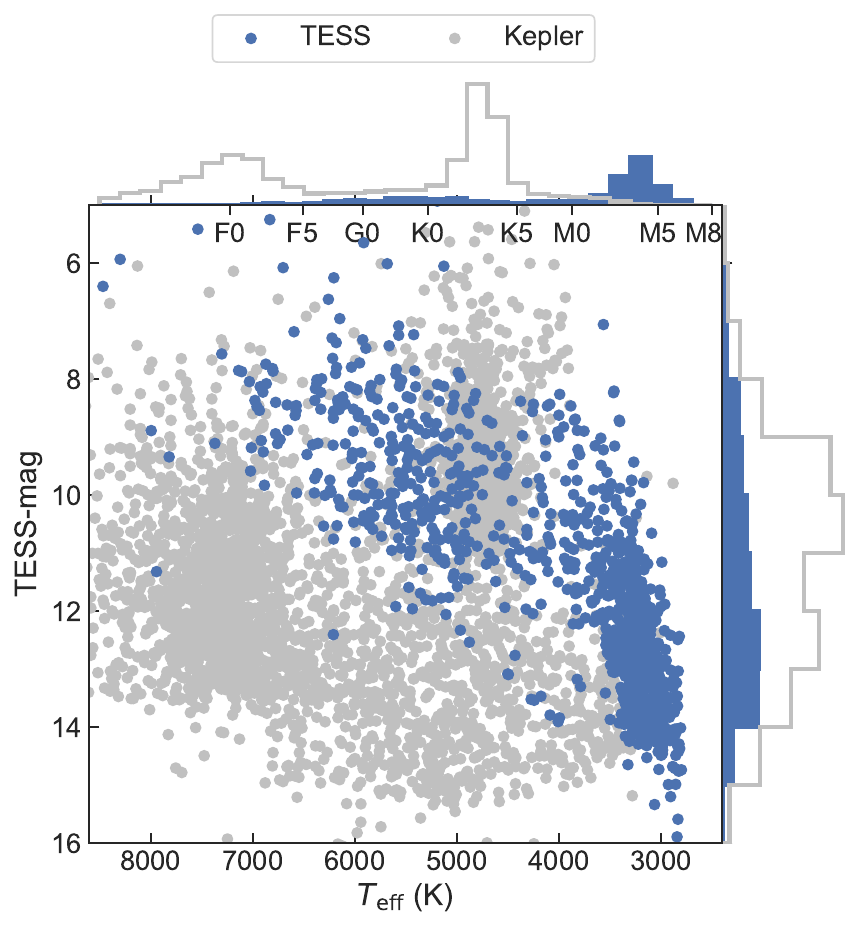}
    \caption{
    TESS explores bright early to late M~dwarfs, expanding the sample size for flare studies in this parameter space. 
    Shown are the effective temperature $T_\mathrm{eff}$ versus TESS magnitude of flaring stars in the TESS sample (blue) and the Kepler flare catalog \citep[grey;][]{Davenport2016}. 
    The top axis indicates stellar types following the classification by \citet{Pecaut2013}. 
    The first two sectors already include {\NconfirmedearlyMdwarfsTESS} early M~dwarfs and {\NconfirmedlateMdwarfsTESS} late M~dwarfs. Out of these, {\NconfirmedearlyMdwarfsTESSbright} and {\NconfirmedlateMdwarfsTESSbright} are brighter than 12th TESS magnitude.
    In addition, TESS detects small flares on F, G, and K dwarfs that are brighter than there average Kepler targets.
    }
    \label{fig:teff_vs_tessmag}
\end{figure}

TESS is designed to survey bright dwarf stars, expanding the number of M~dwarf flares detected in the solar neighborhood (Fig.~\ref{fig:teff_vs_tessmag}).
Our sample includes {\NconfirmedearlyMdwarfsTESSbright} early M~dwarfs and {\NconfirmedlateMdwarfsTESSbright} late M~dwarfs brighter than 12th TESS magnitude. 
These stars can be prime targets for atmospheric characterization and radial velocity follow-up if they are found to host transiting planets.
Flares can provide constraints on the existence of an atmosphere or habitability constraints (see Section~\ref{s:Discussion}).

With 24 more sectors to be observed in the next two years, and flare detection in the full-frame images, the expected TESS yield of flaring M~dwarfs is on the order of $10^4$.
In contrast, the Kepler mission focused on a different sample, namely F, G, and K stars (see Fig.~\ref{fig:teff_vs_tessmag}). The Kepler flare catalog \citep{Davenport2016} collects a total of {\NflarestarsKepler} objects. Of these, {\NflarestarsKeplerxmatched} can be crossmatched with the TIC. The crossmatched list contains {\NflareearlyMdwarfsKepler} flaring early M~dwarfs, and {\NflarelateMdwarfsKepler} flaring late M~dwarf observed with Kepler.
Additional studies with Kepler \citep[e.g.][]{VanDoorsselaere2017}, K2 \citep{Stelzer2016}, MEarth \citep{Mondrik2019} and others can provided an expanded view into the M~dwarf regime, yet still for limited sample sizes. 
Fig.~\ref{fig:teff_vs_tessmag} highlights how TESS enables to explore this parameter space for flare studies in an unprecedented manner.
Moreover, TESS also detects small flares on F, G, and K dwarfs brighter than 12th TESS magnitude, filling in another parameter space which was less explored by the Kepler mission.

\subsection{Flares on fast rotators}
\label{ss:Flares on fast rotators}

Stellar rotation is suggested to be linked to the flaring of a star according to the dynamo theory of magnetic field generation \citep[e.g.][]{Moffatt1978, Parker1979}. The interaction between rotation and convection can cause high magnetic activity \cite[e.g.][]{Browning2008}, whose energy is then released through flares.
Two groups of rotators are observed among low-mass stars: fast rotators ($P_\mathrm{rot}<10$\,d) and slow rotators ($P_\mathrm{rot}>70$\,d) \citep[e.g.][]{Newton2016a,Newton2018}.
Fast rotating M dwarfs ($P_\mathrm{rot}<10$\,d) exhibit a saturated level of activity, while for slower rotators, the ratio of X-ray, H$\alpha$, and Ca H\&K flux to bolometric luminosity declines rapidly \citep{West2015, Astudillo-Defru2017, Newton2017, Wright2018}. The relationship between activity and rotation is typically parameterized in terms of the Rossby number, which seeks to remove the mass dependence by dividing the rotation period by the convective overturn timescale \citep{Noyes1984}. This behavior is consistent across the main sequence from solar-type stars to M dwarfs, and across the fully-convective transition at the lowest stellar masses \citep{Newton2017, Wright2018}. 

We here define two categories: `photometric rotator' and `other'. 
We use the phrase `photometric rotator' synonymous to `a star with a photometrically measured rotation period by TESS or KELT'.
Consequently, rotating stars without detected modulation are not included in this definition.
Our rotation analysis based on TESS data extends to fast rotators with $P_\mathrm{rot}<5$\,d. 
The KELT data covers a limited sample of stars with longer rotation periods.
Hence, the `other' category might still contain many fast rotators with $P_\mathrm{rot}>5$\,d, for which we do not have measured rotation periods.

\begin{figure}[!htbp]
    \centering
    \includegraphics[width=\columnwidth]{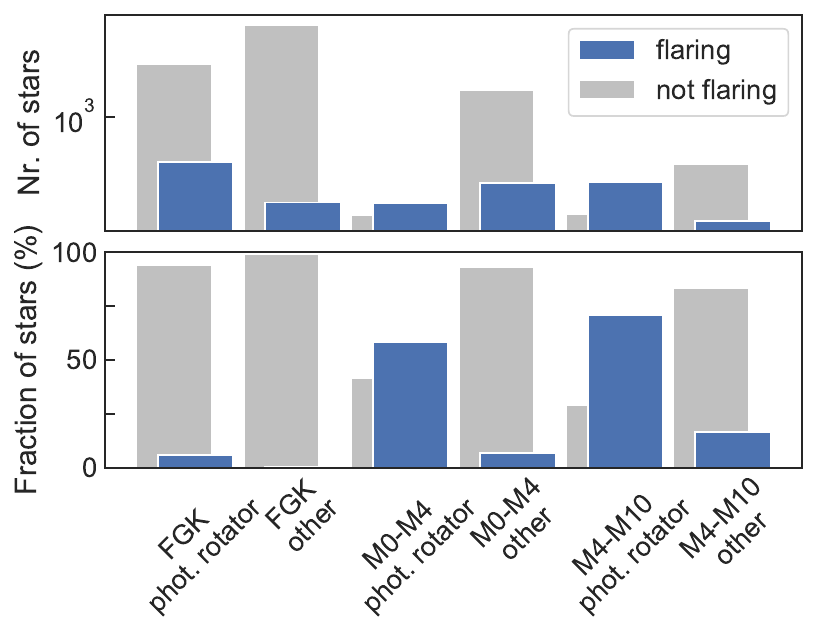}
    \caption{
    Histograms of the number (upper panel) and fraction (lower panel) of flaring stars (blue) compared to ones for which we do not detect flares (grey).
    Bins separate the sample into F, G and K stars, early M~dwarfs of type M0--M4, and late M~dwarfs of type M4--M10; each of these are further grouped by whether or not a rotation period could be photometrically measured.
    Flares are detected for $\sim 60$\% of fast rotating early and late M~dwarfs in the TESS sample. In contrast, only $\sim 10$\% of all other M~dwarfs show detectable flaring. 
    }
    \label{fig:hist_flaring_versus_rotation}
\end{figure}

We find that about $60$\% of fast rotating early and late M~dwarfs in the TESS sample show flaring that is detectable (Fig.~\ref{fig:hist_flaring_versus_rotation}). 
In contrast, only $10$\% of all `other' M~dwarfs have flares.
Among F, G and K stars, less than $5$\% of `photometric rotators' and almost no `others' have flares.
This solidifies past findings from, for example, the Kepler \citep{Davenport2016, VanDoorsselaere2017} and MEarth \citep{Mondrik2019} surveys.

Without fast rotation, there might not be enough energy stored in the magnetic field lines to trigger frequent strong flaring. However, in fully-convective stars the dynamo mechanism is not well-understood. For stars with a radiative zone and a convective zone, the interface between these two zones is what is believed to power the dynamo \citep{Durney93}. Lacking this interface, fully convective M dwarfs (typically with spectral types M4 and later) might generate their magnetic fields through other means \citep[see e.g.][]{Stassun2011}. Nevertheless, the similar behavior in activity and rotation period (measured via H$\alpha$ \citealt{Newton2017} and X-ray \citealt{Wright2018} data) suggests that the manifestations of the magnetic fields in partially and fully convective objects are similar. 

For our sample, it is difficult to disentangle whether the slow rotation impacts both the flare rate and flare amplitude, or just one of them. First, if the flare rate of slow rotators is indeed lower, the 28 days observation span with TESS is not sufficient to detect their flaring.
Second, our detection algorithm is not focused on the detection of the smallest flares; slow rotators could still flare frequently, yet at low amplitudes that are not detected.

\begin{figure}[!htbp]
    \centering
    \includegraphics[width=\columnwidth]{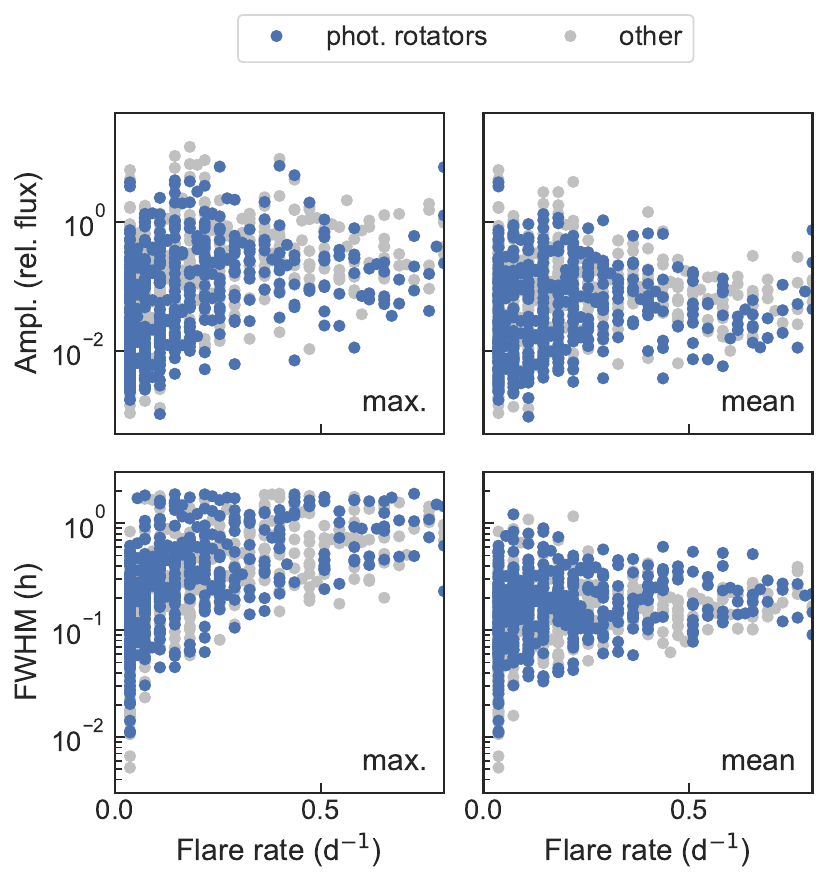}
    \caption{
    The maximum flare amplitude (upper left), mean flare amplitude (upper right), maximum FWHM (lower left), and mean FWHM (lower right) of each star in dependency of the flare rate per day.
    The samples are further separated into stars which have photometrically measured rotation periods (blue) and ones that do not (grey). 
    Stars with higher flare rates have a significantly higher maximum flare amplitude and maximum FWHM.
    There is no significant difference between photometric rotators and other stars.
    }
    \label{fig:amp_and_duration_vs_n_flares_vs_rot}
\end{figure}

Moreover, we find that the maximum flare amplitude and FWHM increase for stars with more frequent flaring (Fig.~\ref{fig:amp_and_duration_vs_n_flares_vs_rot}). 
This seems to be independent of whether a star is a photometric rotator or not.
Furthermore, stars with higher flare rates show similar mean values of their flares amplitudes and durations.

\begin{figure}[!htbp]
    \centering
    \includegraphics[width=\columnwidth]{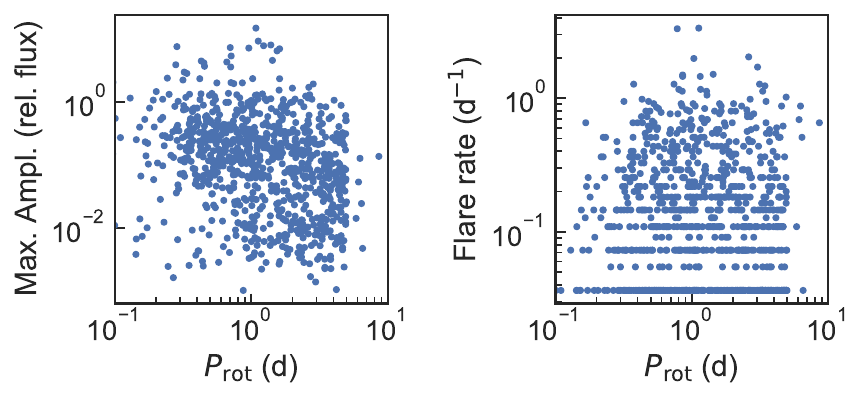}
    \caption{
    The maximum flare amplitude (left panel) and number of flares per day (right panel) are studied as a function of the rotation periods $P_\mathrm{rot}$ measured by TESS and KELT.
    There is no significant correlation between photometric fast rotators and the amplitude of their superflares, albeit we note that our sample is mostly limited to $P_\mathrm{rot}<1$.
    There are signs that the fastest rotators ($P_\mathrm{rot}<0.3$\,d) do not flare as frequently as the rotators with $P_\mathrm{rot}>0.3$\,d. 
    However, this remains to be re-examined with a larger sample size.
    }
    \label{fig:amp_and_n_flares_vs_rot}
\end{figure}

We do not find a correlation between stars' rotation periods and their flare amplitudes (Fig.~\ref{fig:amp_and_n_flares_vs_rot}).
Large flares seem to be as likely to occur on photometric fast rotators as on the other stars in our sample.
This is in agreement with previous findings \citep[e.g.][]{Maehara2012}, and can be explained if the magnetic field energy is stored near a star spot. It was shown for different star types, that in this scenario the rotation period does not influence the maximum energy \citep[e.g.][]{Rodono1986}.

Our results suggest a link between the flare frequency and the rotation period for the fastest photometric rotators (Fig.~\ref{fig:amp_and_n_flares_vs_rot}). Stars with $P<0.3$\,d  reach only half the maximum flare frequency as those with $P>0.3$\,d. 
However, this could also be sampling issue; there are fewer photometric rotators with $P<0.3$\,d and the distribution of the flare frequency, peaking at $\sim0.1$ flares per day, shows a long tail until $\sim0.5$ flares per day.

For $P>1$\,d, the frequency of superflares (flares with energies $>10^{33}$\,erg) was reported to decrease with period \citep[e.g.][]{Maehara2012}. For our sample, however, the available rotation period information is too sparse to study this link for rotation periods $P>1$\,d.

\subsection{99 stars show flares with flux increases of multiple magnitudes}
\label{ss:XX stars show superflares with flux increases of multiple magnitudes}

\citet{Chang2018} conducted a study of flaring M~dwarf stars in the Kepler sample. They found eight flares that increase the stellar brightness at peak by a factor of two or more. 
The authors define these events as `hyperflares' (we do not follow this nomenclature here).
The strongest flare they detected has a peak luminosity 5 times the quiescent flux.

The TESS sample profits from the mission's observing strategy, allowing us to identify {\Nconfirmedhyperflarestars} flaring stars with {\Nconfirmedhyperflares} individual flares that fulfill the \citet{Chang2018} criterion of at least doubling the stellar brightness.
A list of these targets can be created from Table~\ref{tab:catalog_per_flare} by sorting by the amplitude column.

\subsection{The largest superflare and the most energetic superflare in the TESS sample}
\label{ss:The largest superflare in the TESS sample}

\begin{figure}[!htbp]
    \centering
    \includegraphics[width=\columnwidth]{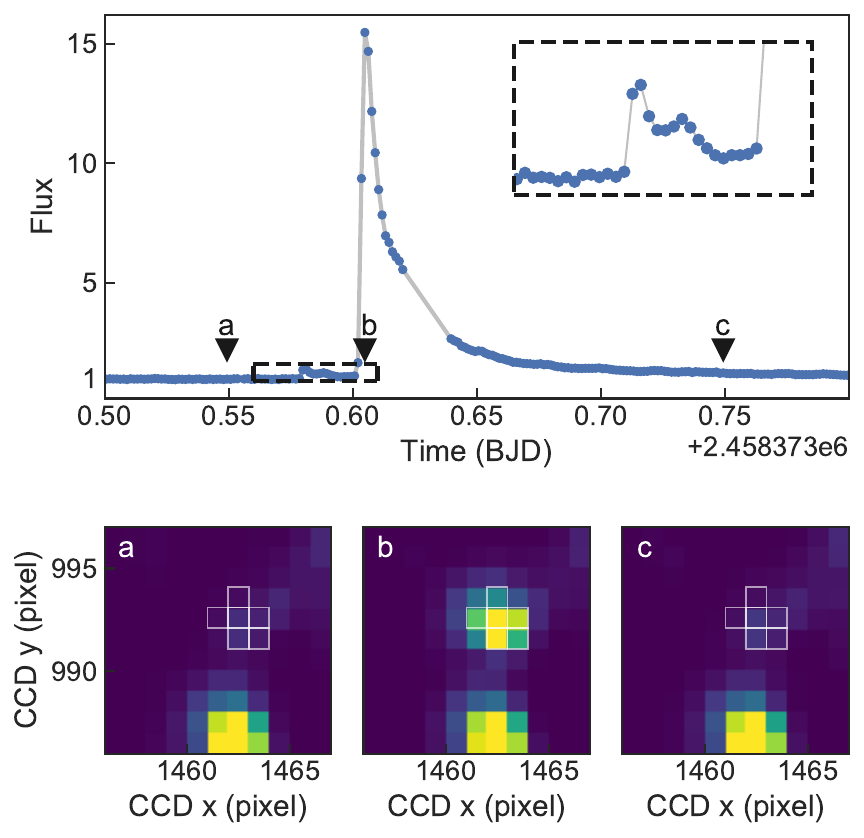}
    \caption{The largest flare in our sample is on {\TICone}, an M4.5V dwarf star ({\TIConeteff}\,K) with TESS-mag {\TIConetessmag}.
    Upper panel: TESS PDC-SAP lightcurve.
    Lower panel: TESS target pixel images at the times before the flare (a), during the flare peak (b), and after the flare (c), as indicated in the upper panel.
    Our best fit model shows that during the peak the star increases its brightness by a factor of {\TIConeamplmax}. This superflare is preceded by a series of smaller flares (inlay). We inspect the target pixel files for this outburst, and confirm that the flare originates from the target.}
    \label{fig:16x_flare}
\end{figure}

{\TICone} (2MASS J06270005-5622041), an M4.5V dwarf star ({\TIConeteff}\,K) with TESS-mag {\TIConetessmag}, shows the largest flare amplitude in our sample (Fig.~\ref{fig:16x_flare}). Over the course of 1~h, the star increases its optical brightness by a factor of {\TIConeamplmax}, releasing a bolometric energy of {\TIConebolenergymax}\,erg.
The superflare is preceded by a series of smaller flares. 
We inspect the individual target pixel files for this outburst using the \texttt{lightkurve} \citep{lightkurve} module.
This confirms that the flare is indeed on TIC~260506296.

In contrast, the most energetic superflare in our sample reaches {\bolenergymax}\,erg. It is found on {\TICtwo} (2MASS J23211550-2659121), a G-type giant star ({\TICtwoteff}\,K, {\TICtworadius}~$\mathrm{R_\odot}$) with TESS-mag {\TICtwotessmag}. 
Such energetic flares on giant stars are common \citep[e.g.][]{Balona2015, VanDoorsselaere2017}, which is surprising, as the evolution off the main sequence should decrease their magnetic field \citep{Simon1989}. Possible explanations could be that the giant's surface magnetic field is conserved, or that the flare actually originates from an unresolved dwarf star companion. 
Notably, {\TICtwo}'s enormous release of energy only leads to a flare amplitude of {\TICtwoamplmax}.
This is a direct consequence of the stellar type, namely effective temperature and radius, following Eqs.~\ref{eq:flare_energy_start}--\ref{eq:flare_energy_end}.
In comparison, the energy output for the M~dwarf TIC~260506296 is significantly lower, despite its much larger amplitude; this is due to the M~dwarf's small radius and lower effective temperature.

\subsection{Flare frequency distributions}
\label{ss:Flare energies versus flare rate}

\begin{figure*}[!htbp]
    \centering
    \includegraphics[width=\textwidth]{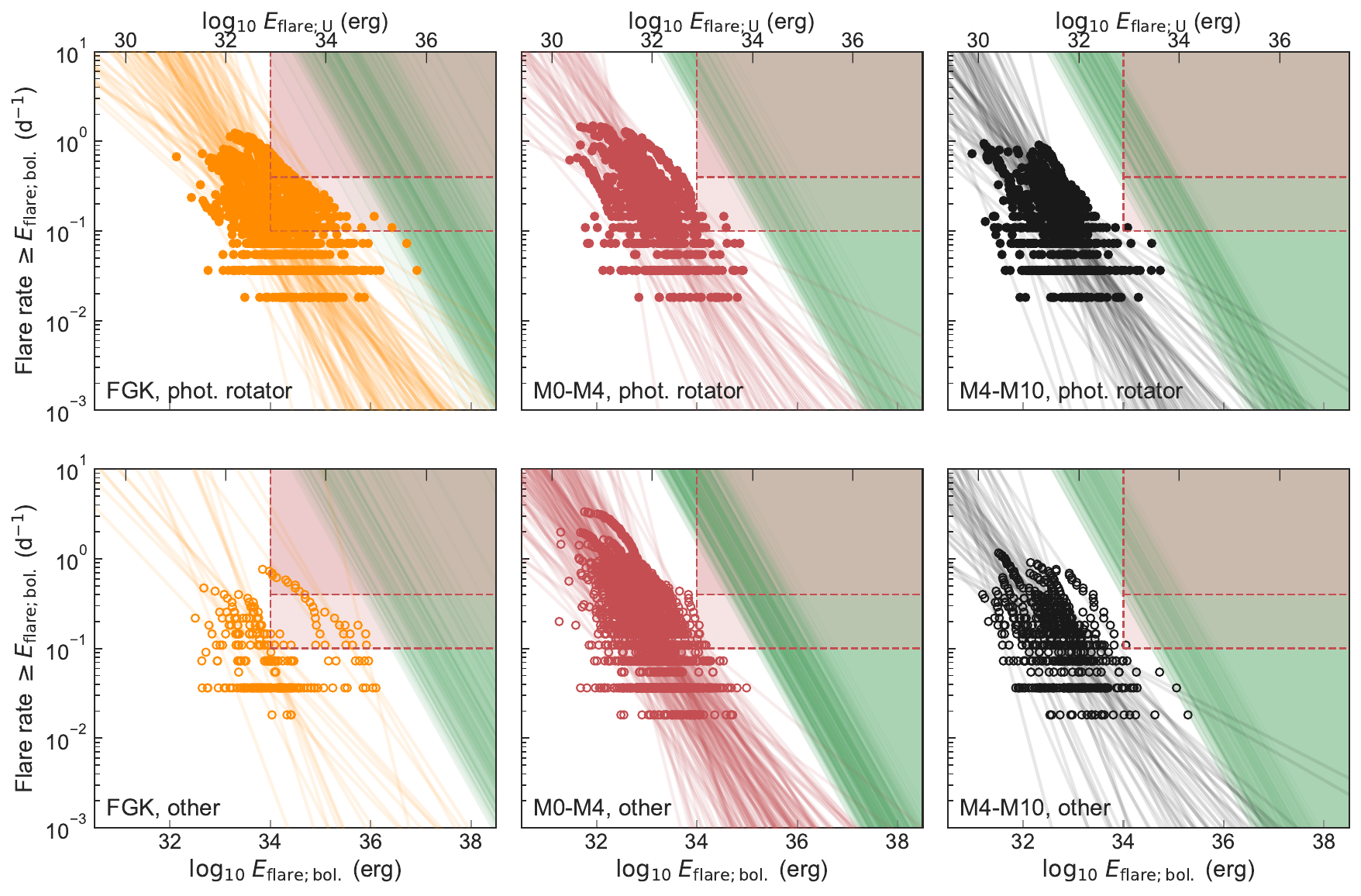}
    \caption{Flare frequency distributions (FFD) in the context of prebiotic chemistry (green area) and ozone sterilization (red areas).
    The x-axis shows the flare energy, as bolometric energy $E_\mathrm{flare; bol.}$ on the lower ticks, and U-band energy $E_\mathrm{flare; U}$ on the upper ticks. 
    The y-axis shows the cumulative rate of flares per day, i.e. how often a flare with at least a certain energy appears.   
    Different panels show F, G and K stars (orange), early M dwarfs (red) and late M dwarfs (black), separated into photometric rotators (filled circle) and others (unfilled circle).
    Solid lines are linear fits to the double-logarithmic FFD of each star, extrapolating into regimes that could not be observed.
    The green area denotes the minimum flare rate and energy required to trigger prebiotic chemistry on a potential exoplanet \citep[expanded from][see Section~\ref{ss:Which flares deliver enough energy for prebiotic chemistry?}]{Rimmer2018}. 
    The different green shadings show each threshold for each star, which depends on the stellar radius and effective temperature (Eq.~\ref{eq:preb_chem}).
    In the red shaded region derived from \citet{Tilley2019}, intense flares are frequent enough that ozone layers cannot survive, and planet surfaces may be sterile (see Section~\ref{ss:Habitability around TIC 260506296}).
    We mark two ozone sterilization regions: a permissive threshold for flare rates $\geq 0.1$ per day (lighter red area), and a conservative threshold for flare rates $\geq 0.4$ per day (darker red area).
    {\Nconfirmedprebchem} stars, including {\NconfirmedprebchemearlyM} early M~dwarfs and {\NconfirmedprebchemlateM} late M~dwarfs, in the TESS sample fulfill the criteria of prebiotic chemistry. On the other hand, potential exoplanets around {\Nconfirmedozonedeplperm} stars, including only {\NconfirmedozonedeplpermM} M~dwarfs, might suffer from ozone depletion.
    }
    \label{fig:FFD_vs_preb_chem}
\end{figure*}

We study the flare energies and flare rates as a function of the stellar type and rotation period (Fig.~\ref{fig:FFD_vs_preb_chem}).
This is commonly denoted as the flare frequency distribution \citep[FFD; e.g.][]{Gershberg1972, Lacy1976, Hawley2014}.
The FFD shows the cumulative rate of flares per day, i.e. how often a flare of a certain energy or higher is detected. 

Using the TIC information, we separate F, G, and K stars from early M~dwarfs (M0--M4; 3905K--3200K) and late M~dwarfs (M4--M10; 3200--2285K). These are further divided into stars with photometrically measured rotation modulation (`phot. rotator') and ones without (`other'; see Section~\ref{ss:Flares on fast rotators}).
Each star's FFD is fit with a line in the log-log space to extrapolate the trend, following $\log_{10} (\mathrm{flare~rate}) = \alpha \log_{10} (E_\mathrm{bol.}) + \beta$. The best fit parameters $\alpha$ and $\beta$ are included in Table~\ref{tab:catalog_per_flare}.

We find no significant difference between stars with and without detected rotation modulation. This is somewhat surprising given that fast rotating M~dwarfs are suggested to flare more frequently.
Moreover, we find that the average flare energy per star depends on the effective temperature and radius of the star, which is a direct consequence of Eqs.~\ref{eq:flare_energy_start}--\ref{eq:flare_energy_end}. 
For F, G and K stars, this effect in addition with their lower contrast (see Section~\ref{ss:TESS explores a new flare sample: bright M dwarfs}), is why they generally show higher flare energies than early M dwarfs and late M dwarfs.

\section{Discussion}
\label{s:Discussion}

\subsection{Which flares deliver necessary energy to trigger prebiotic chemistry?}
\label{ss:Which flares deliver enough energy for prebiotic chemistry?}

We combine the measured flare energies and rates with the laboratory study conducted by \citet{Rimmer2018}, who delineate `abiogenesis zones'. 
These are zones outside of which a specific prebiotic chemical scenario cannot succeed around main sequence stars. 
This considers ribonucleotides, required for ribonucleic acid (RNA) synthesis, as a starting point for prebiotic chemistry \citep[e.g.][]{Sutherland2015}.
The authors consider the competition between reactions that produce ribonucleotides in the presence of UV light (200-280 nm), with other bimolecular reactions that produce inert adducts with no prebiotic interest. 
They calculate the UV light from a flare multiplied by the flare frequency, and compare it to the lifetime of prebiotic intermediates required for ribonucleotide formation
\citep[e.g.][]{Xu2018}.
\citet{Rimmer2018} develop equations for a planet at a fixed distance from its host star. Here, we adjust the star-planet distance, $a$, to keep the planet at a distance from its host star so that the flux is the same as 1 Earth flux \citep[as defined by][]{Kane2016}. The flare frequency needed to drive the prebiotic chemistry, $\nu$, is a function of the flare's U-band energy $E_U$, the stellar radius, $R_*$, and the stellar temperature, $T_*$. Then, $\nu$ can be derived following:
\begin{equation}
\nu \geq 25.5 \, {\rm day^{-1}} \, \Bigg(\dfrac{10^{34} \, \rm{erg}}{E_U}\Bigg)\Bigg(\dfrac{R_*}{R_{\odot}}\Bigg)^{\!\!2}\Bigg(\dfrac{T_*}{T_{\odot}}\Bigg)^{\!\!4}.
\label{eq:preb_chem}
\end{equation}

To apply this equation to the bolometric energies we derived, we first calculate the flares' U-band energy from the bolometric energy. This uses the U-band spectral response function and assumes a 9000\,K blackbody for the flare. We find that 7.6\% of the flare's bolometric energy falls into the U-band, leading to $E_\mathrm{U} \approx 7.6\% E_\mathrm{bol}$.

Flare frequencies satisfying Eq.~\ref{eq:preb_chem} could drive the synthesis of pyrimidine ribonucleotides, the building blocks of RNA, from hydrogen cyanide and bisulfite in liquid water. These flare rates establish a \textit{necessary} condition for the origin of life's building blocks according to the scenario proposed by \citet{Xu2018}. Having the necessary UV light from stellar activity is \textbf{not} a \textit{sufficient} condition for this scenario. Liquid water, hydrogen cyanide, bisulfite, and other feedstock molecules are also necessary.

Fig.~\ref{fig:FFD_vs_preb_chem} displays the flare rates of the stars in our sample, compared to this inequality.
We find that {\Nconfirmedprebchem} stars have enough flaring (at their current age) to deliver the UV energy at a rate that could trigger prebiotic chemistry on a potential exoplanet.
These include {\NconfirmedprebchemearlyM} early M~dwarfs and {\NconfirmedprebchemlateM} late M~dwarfs.
Out of these, {\NconfirmedprebchemearlyMbright} early M~dwarfs are brighter than 12th TESS magnitude.
A list of the targets whose flaring have this potential can be created from Table~\ref{tab:catalog_per_flare} by sorting by the `prebiotic chemistry' column.

It is important to note that most of the stars in our sample are likely old (albeit we currently have limited knowledge of their exact ages). The onset of prebiotic chemistry likely occurred during earlier stages of exo-solar systems, when the stars were younger. Given that young M~dwarfs were more active \citep{West2008} and thus assuming that their flaring was more frequent and energetic in the past, our results provide a lower limit. If a star's FFD passes this cutoff now, it will also have done so in a younger stage.

\subsection{The impact of coronal mass ejections}
\label{ss:The impact of coronal mass ejections}

A coronal mass ejection (CME) often follows a stellar flare, and can have a substantial impact on potential exoplanets. A CME is a large release of plasma and the connected magnetic field from the stellar corona. 

To estimate the impact of CMEs, we apply the empirical relationship between flare energy and CME mass found by \citet{Aarnio2012}. This has been calibrated against observations of flares and CMEs from the Sun, as well as a sample of well-studied, very active pre-main-sequence stars. The callibration encompasses flare energies from $\sim$10$^{28}$\,erg up to $\sim$10$^{38}$\,erg---comparable to the strongest flares included in this study---and which correspond to CME masses of up to $\sim$10$^{22}$~g \citep{Aarnio2011,Aarnio2012}. 

This empirical relation, however, is applicable to the flare energy that is emitted in X-rays, whereas
in this work we determined the flare energy that is emitted bolometrically (Section~\ref{ss:Measuring the flare rate}). 
To apply it to our data, we estimate the X-ray flare energy as 1\% of the bolometric energy, which is based on findings for the strongest solar flares. These have X-ray fluxes of about $10^{-4}$ W m$^{-2}$ \citep{Aarnio2011}, corresponding to X-ray energies of $10^{29}$\,erg \citep{Aarnio2012} and to bolometric energies of $10^{31}$ erg \citep{Maehara2015}. Hence, for the same flare strength, the energy emitted bolometrically is typically larger by a factor of 100 \citep[see also][]{Osten2015}.

Thus we adapt the following equation from \citet{Aarnio2012} for the CME mass, $M_\mathrm{CME}$, in terms of the bolometric flare energy:
\begin{align}
M_\mathrm{CME} = (2.7 \pm 1.2) \times \left(\frac{E_\mathrm{flare;bol}}{100}\right)^{(0.63 \pm 0.04)}
\label{eq:CME}
\end{align}
All values are in cgs units.

From Eq.~\ref{eq:CME}, we find a median CME mass of {\MCMEmedian}~g, ranging from a minimum of {\MCMEmin}~g to {\MCMEmax}~g
for the most massive CME. This could have accompanied the strongest flare in our sample, which was found on the giant star {\TICtwo} ({\TICtwoteff}\,K, {\TICtworadius}~$\mathrm{R_\odot}$).
All estimates are also given in Tables~\ref{tab:catalog_per_flare} and \ref{tab:catalog_per_star}.

CME events mainly happen along a given direction and, as such, are non-isotrop events. 
Their impact depends among other factors on the semi-major axis and inclination of the exoplanet's orbit \citep{Kay2016}.
As exoplanets' habitable zones around M~dwarfs are close to their star, the probability of being hit by a CME is higher than for their counterparts around Sun-like stars.
Even though the overall chance of a strong CME hitting an exoplanet is relatively low, it is interesting to consider the potential impact on habitability. CMEs impinging on a nitrogen-rich atmosphere can be useful for prebiotic chemistry via the efficient production of hydrogen cyanide, and possibly for atmospheric warming, via the production of nitrous oxide \citep{Airapetian2016}. On the other hand, CMEs can contribute significantly to the potential loss or transformation of exoplanet atmospheres through ion pick up \citep{Lammer2007,Cohen2015}.
For M~dwarfs, the exoplanet's magnetic field has to be up to hundreds of Gauss strong to shield from atmospheric loss through CMEs, making them a more detrimental factor than stellar winds \citep{Kay2016}.
Very energetic particles from the CME may produce secondary particles that can reach the surface, changing mutation rates for life there \citep{Smith2004}.

\subsection{Ozone depletion}
\label{ss:Ozone depletion}

The radiation from stellar proton events (SPEs) could deleteriously impact the habitability of exoplanets.
However, \cite{Atri2017} find that while energetic SPEs could trigger extinction events among complex life, SPEs are unlikely to lead surface sterilization. In particular, exoplanets with magnetic fields and/or substantial atmospheres are sufficiently shielded from SPEs. 

The impact of protons and photons from M~dwarf stellar flares on the atmosphere of a modern-Earth analog exoplanet has been modeled by \citet{Tilley2019} \citep[see also e.g.][]{Youngblood2017}. They find that the impact of flares of energy $10^{34}$ ergs at a frequency of $1~\text{month}^{-1}$ or greater will result in the removal of $99.99\%$ or more of the ozone layer, incidentally admitting potentially-sterilizing doses of UV radiation to the planet surface. \citet{Tilley2019} estimate that, conservatively, $0.083$ of M~dwarf flares will strike a habitable-zone exoplanet, meaning that terrestrial-type ozone layers should not persist on stars with flares of $\geq 10^{34}$ erg at a frequency of $\geq0.4~\text{day}^{-1}$. 
A more permissive limit could be drawn at $\geq0.1~\text{day}^{-1}$. These parameter spaces are marked in red in Figure~\ref{fig:FFD_vs_preb_chem}. 

Few of the stars we consider in this study fall into this parameter space.
For the permissive limit of $0.1~\text{day}^{-1}$, potential exoplanets around {\Nconfirmedozonedeplperm} stars, including {\NconfirmedozonedeplpermM} M~dwarfs, might suffer from ozone depletion.
For the conservative limit of $0.4~\text{day}^{-1}$, {\Nconfirmedozonedeplcons} stars could be affected.
However, it is possible that more frequent but smaller flares and/or less frequent but more energetic flares may also remove ozone layers. Further modelling work is required to rule on this possibility. We note that even in the absence of an ozone layer, life can survive in the ocean or the subsurface, meaning that flares may not be a strict barrier to habitability. 

From a different point of view, ozone has been proposed as a biosignature \citep{Segura2003}. Exoplanets subject to sufficiently intense flares would not be able to develop ozone layers, leading to a potential `false negative' scenario for ozone. Other biosignatures may be similarly affected. The FFDs we derive in this paper will enable detailed modelling of the atmospheric states of planets orbiting these stars, and will enable the assessment of false positive and false negative scenarios for biosignature search for these objects.

\subsection{Additional constrains for habitability}
\label{ss:Additional constrains for habitability}

We discussed selected aspects of how flare events can influence an exoplanet's habitability: how flares could trigger prebiotic chemistry (Section~\ref{ss:Which flares deliver enough energy for prebiotic chemistry?}), the impact of coronal mass ejections accompanied with flares (Section~\ref{ss:The impact of coronal mass ejections}), and ozone depletion (Section~\ref{ss:Ozone depletion}).

A more comprehensive view of habitability should also consider the following factors: the exoplanet's orbit (semi-major axis and eccentricity); the stellar type; stellar companions; the exoplanet's atmospheric composition or absence of an atmosphere; the exoplanet's surface; potential sub-surface habitability; and many other factors \citep[see e.g.][]{Seager2013, Barnes2015, Shields2016, Meadows2018}.

\subsection{On the habitability around TIC 260506296\\ and other superflare M~dwarfs}
\label{ss:Habitability around TIC 260506296}

One of the most interesting targets for habitable exoplanet search will be M~dwarfs whose FFDs fill the lower right corner of the panels in Fig.~\ref{fig:FFD_vs_preb_chem}. In these areas, the flaring is energetic enough to trigger prebiotic chemistry (green area), but not with such a large frequency that would lead to ozone depletion (red areas). This is an particularly interesting parameter space for M~dwarfs, which can soon be explored further with continuous one-year observations in the TESS continuous viewing zone.
Note, however, that we here assume the FFDs can be extrapolated in this regime; findings from Kepler suggest that this might not always be the case for M~dwarfs \citep[e.g.][]{Davenport2016}.

What can we say about TIC~260506296, which shows the largest superflare in the sample? Its flux increases by a factor of {\TIConeamplmax} in the TESS band over the course of 1\,h, releasing a bolometric energy of {\TIConebolenergymax} erg (see Section~\ref{ss:The largest superflare in the TESS sample}).
In contrast, the Carrington event on our Sun (G2V; 5772\,K) released $10^{32}$\,erg. Yet, the flare energy of these events on M~dwarfs depend on the quiescent flux. Taking the radius and temperature differences between M~dwarfs and the Sun into account (Eqs.~\ref{eq:flare_energy_start}--\ref{eq:flare_energy_end}), it becomes clear how much more extreme flares are on these prime targets for exoplanet missions.

The superflare on {\TICone} is comparable to, yet smaller than, the largest superflare reported for Proxima Centauri by \citet{Howard2018}. The Proxima Centauri superflare temporarily increased the star's brightness by a factor of at least 38, releasing $10^{33.5}$\,erg.
Interestingly, {\TICone} (M4.5V, {\TIConeteff}\,K) and Proxima Centauri (M5.5V, 3050\,K) are comparable stars. 
Such large stellar flares have been suggested to significantly impact the habitability of putative planetary companions (e.g., \citealt{Howard2018}).

We find that {\TICone} indeed provides substantial flaring to trigger prebiotic chemistry on nearby worlds (see Tables~\ref{tab:catalog_per_flare} and \ref{tab:catalog_per_star}).
Its superflare was potentially accompanied by a CME event of $10^{23}$\,g, and the star's FFD does not fall into the regime of ozone depletion by stellar proton events.
All of this is opening interesting avenues for future studies on this potential exoplanet host, as well as other comparable M~dwarfs.

\subsection{Four TESS exoplanet candidates orbiting flaring M~dwarfs}
\label{ss:Can we find planets around these stars?}
Whether we can actually detect exoplanets around such strongly flaring stars is a separate matter. A large fraction of them are fast rotators with high flare frequencies, which can inhibit the detection of small planets \citep[see e.g.][]{Berta2012, Kipping2017}. Additionally, planetary mass measurements can be hindered if the stellar rotation and planetary orbital periods are similar or harmonics. This is particularly a problem for habitable-zone planets around field-aged early M dwarfs \citep[][]{Newton2016a, Vanderburg2016}.

We do not find any known exoplanets among our flare stars when cross-matching Table~\ref{tab:catalog_per_star} with the known exoplanets lists provided by Stephen Kane\footnote{\url{heasarc.gsfc.nasa.gov/docs/tess/data/approved-programs/G011183.txt}, online 28 Dec 2018} and John Southworth\footnote{\url{heasarc.gsfc.nasa.gov/docs/tess/data/approved-programs/G011112.txt}, online 28 Dec 2018} for the TESS Guest Observer program.

We do, however, find four matches among the current TESS alerts\footnote{\url{archive.stsci.edu/prepds/tess-data-alerts/}, online 28 Dec 2018}, which could potentially be exoplanets transiting flaring stars from our sample:
\begin{itemize}
\item TIC~32090583 is an M~dwarf with a $4.84\,\mathrm{R_\oplus}$ exoplanet candidate on a 0.438\,day orbit.
\item TIC~70797900 is an early M~dwarf hosting a single-transit candidate with unknown period. However, its large candidate radius of $17.06\,\mathrm{R_\oplus}$ and V-shaped transit could indicate an eclipsing binary scenario.
\item TIC~206609630 is potentially an M~dwarf with a $5.51\,\mathrm{R_\oplus}$ exoplanet candidate on a 0.335\,day orbit (note that it has a \textit{Gaia} source duplication flag).
\item TIC~272086159 is a mid M~dwarf with a $9.77\,\mathrm{R_\oplus}$ exoplanet candidate transiting every 16.156\,days.
\end{itemize}

These exo-solar systems, if confirmed, could allow interesting case studies. However, the four candidates themselves likely will have thick gas envelopes, increasing the temperature beneath the atmosphere to a level which is too high for liquid water and a solid surface. Nevertheless, potential planetary companions or exomoons \citep[see e.g.][]{Teachey2018} might provide the necessary conditions.

\section{Conclusion}
\label{s:Conclusion}

We detect and analyze stellar flares in the short cadence (2~minute) lightcurves from the first TESS data release.
To do so, we develop a flare detection pipeline, whose candidates we visually inspect to create a vetted flare candidate list.
We then apply our newly developed \texttt{allesfitter} software to fit the profiles of each flare with different models, ranging from pure noise to complex flare sequences. 
Using Nested Sampling, we compute the Bayesian evidence of each model. This allows us to robustly select the favoured scenario.

We find {\Nconfirmeduniquetargets} flaring stars in the first two TESS sectors, with a total count of {\Nconfirmedflares} flares.
The largest amplitude flare appears on the M~dwarf {\TICone} and increases the brightness by a factor of {\TIConeamplmax}.
The flare with the highest energy output is on the G-type giant star {\TICtwo}, releasing {\TICtwobolenergymax}\,erg and could be accompanied by a coronal mass ejection (CME) of {\TICtwoMCME}\,g.

Among all observed stars, flares appear on 30\% of mid to late M~dwarfs, on 5\% of early M~dwarfs, and on less than 1\% of F, G and K stars.
{\NconfirmedMdwarfsTESS} of all flaring stars are early to late M~dwarfs, highlighting that TESS explores an important parameter space for flare studies.
In total, we find {\NconfirmedearlyMdwarfsTESS} early M~dwarfs and {\NconfirmedlateMdwarfsTESS} late M~dwarfs.

Moreover, we investigate flaring as a function of photometrically measured stellar rotation periods (`photometric rotator'). 
60\% of fast rotating M~dwarfs are flare stars. Of the M~dwarfs without detected rotation periods, only 10\% flare.
We further find that star's with higher flare rates also have an increased maximum flare amplitude and FWHM. 
Photometric rotators and other stars show comparable flare rates, amplitudes, and energies; however, we note that our rotation information mainly covers the regime of $P_\mathrm{rot}<5$\,d. 
Among photometric rotators, there is a tentative decrease of the flare rate for $P_\mathrm{rot}<0.3$\,d.

We analyze the flare frequency distributions (FFDs) in the context of prebiotic chemistry, CMEs and ozone depletion. 
On the one hand, flares have been suggested to deliver the required ultraviolet energy to trigger biogenesis on exoplanets. We find {\Nconfirmedprebchem} stars, including {\NconfirmedprebchemM} M~dwarfs, which could fulfill the criteria of a minimum flare rate and energy. 
Nevertheless, most stars do not seem to be able to provide the necessary ultraviolet energy through their flares alone. 
On the other hand, CMEs and stellar proton events (SPEs) associated with flares could further impact existing life, with SPEs potentially causing ozone depletion for exoplanet's atmospheres. We find that potential exoplanets around up to {\Nconfirmedozonedeplperm} stars might suffer from this effect. Noteably, these are mostly F, G, and K type stars, while only up to {\Nconfirmedozonedeplperm} M~dwarfs seem affected. This can negatively impact habitability and the search for biosignatures.

Four flaring M~dwarfs host exoplanet candidates from recent TESS alerts: TIC 32090583, TIC 70797900, TIC 206609630, and TIC 272086159. Together with other M~dwarfs, such as the superflaring {\TICone}, these systems can open an interesting avenue for future studies of habitability.

It is important to note that these findings alone do not allow to rule on the possibility of life on potential exoplanets. Our study provides novel methods and insights to derive an overview. Yet, when addressing complex topics such as exoplanet habitability, any interesting individual system deserves a detailed study to consider a variety of interrelated factors. Fortunately, TESS will continue delivering these prime targets.
Those in the mission's continuous viewing zone in particular will provide reliable statistics on the largest, and potentially rarest, superflares.

\section*{Acknowledgments}
We thank Ward Howard for engaging discussions about stellar flares.
Funding for the TESS mission is provided by NASA's Science Mission directorate. 
This paper includes data collected by the TESS mission, which are publicly available from the Mikulski Archive for Space Telescopes (MAST).
MNG acknowledges support from MIT's Kavli Institute as a Torres postdoctoral fellow.
TD acknowledges support from MIT's Kavli Institute as a Kavli postdoctoral fellow.
PBR acknowledges support from the Simons Foundation [SCOL awards 599634].
This work was supported in part by grants from the Simons Foundation (SCOL  grant \# 495062 to S.R). MHK acknowledges Allan R. Schmitt and Troy Winarski for making their light curve softwares freely available.
This research has made use of IMCCE's SkyBoT VO tool.

\textit{Facilities}:
{\TESS} \citep{Ricker2014}

\textit{Software}:
{\scshape python} \citep{Rossum1995},
{\scshape numpy} \citep{vanderWalt2011},
{\scshape scipy} \citep{Jones2001},
{\scshape matplotlib} \citep{Hunter2007},
{\scshape tqdm} (\url{doi:10.5281/zenodo.1468033}),
{\scshape seaborn} (\url{seaborn.pydata.org/index.html}),
{\scshape allesfitter} \citep{allesfitter-paper, allesfitter-code},
{\scshape ellc} \citep{Maxted2016},
{\scshape aflare} \citep{Davenport2014},
{\scshape dynesty} \citep{Speagle2020},
{\scshape emcee} \citep{Foreman-Mackey2013},
{\scshape celerite} \citep{Foreman-Mackey2017},
{\scshape corner} \citep{Foreman-Mackey2016}.

\begin{deluxetable*}{lllllllllll}
\tablecaption{Catalog of all individual flares found in TESS sectors 1 and 2.\label{tab:catalog_per_flare}}
\tablehead{
\colhead{TIC ID} & 
\colhead{sector} & 
\colhead{outburst} & 
\colhead{flare} & 
\colhead{$t_\mathrm{peak}$} &
\colhead{Amp.} & 
\colhead{FWHM} & 
\colhead{$E_\mathrm{bol}$} & 
\colhead{$M_\mathrm{CME}$} &
\colhead{...} \\
\colhead{} & 
\colhead{} & 
\colhead{} & 
\colhead{} & 
\colhead{(BJD)} &
\colhead{(rel. flux)} & 
\colhead{(d)} & 
\colhead{(erg)} & 
\colhead{(g)} &
\colhead{...}
} 
\startdata
\multicolumn{11}{c}{...} \\
167602025 & 2      & 2        & 1     & 2458373.524 & 0.008 & 0.006    & 8.15e+33    & 3.42e+20 & ... \\
167695269 & 1      & 1        & 1     & 2458326.726 & 0.093 & 0.002    & 7.62e+32    & 8.22e+19 & ... \\
167695269 & 1      & 1        & 1     & 2458336.636 & 0.157 & 0.004    & 2.18e+33    & 1.71e+20 & ... \\
167695269 & 1      & 2        & 1     & 2458343.026 & 0.512 & 0.002    & 4.25e+33    & 2.74e+20 & ... \\
167696018 & 1      & 1        & 1     & 2458325.491 & 0.010 & 0.010    & 5.43e+33    & 3.23e+20 & ... \\
167696018 & 1      & 2        & 1     & 2458328.993 & 0.013 & 0.019    & 2.18e+34    & 5.24e+20 & ... \\
\multicolumn{11}{c}{...} \\
\hline
\multicolumn{11}{c}{
\begin{minipage}{0.9\textwidth}
With one row for each flare, the following values are listed:
the \textit{TIC ID}; 
the TESS \textit{sector};
the \textit{outburst} number;
the \textit{flare} number;
the posterior median for the peak time (\textit{$t_\mathrm{peak}$}), amplitude (\textit{Amp.}) and full-width at half-maximum (\textit{FWHM});
the bolometric energy of the flare (\textit{$E_\mathrm{bol.}$});
and the possible mass of a coronal mass ejection following the flare (\textit{$M_\mathrm{CME}$}; see Section~\ref{ss:The impact of coronal mass ejections}).
The machine-readable version is available in the online journal. It additionally contains lower and upper limits where applicable. For ease-of-use, it also includes a copy of the per-star columns shown in Table~\ref{tab:catalog_per_star} (see below).
\end{minipage}}\\
\enddata
\end{deluxetable*}

\begin{longrotatetable}
\begin{deluxetable*}{llllllllllllllllll}
\tablecaption{Catalog of all flaring stars found in TESS sectors 1 and 2.\label{tab:catalog_per_star}}
\tablewidth{700pt}
\tabletypesize{\footnotesize}
\tablehead{
TIC ID    & 
$N_\mathrm{sec.}$ & 
$N_\mathrm{out.}$ & 
$N_\mathrm{fla.}$ & 
Amp. & 
... &
$M_\mathrm{CME}$ &
TESS- & 
$T_\mathrm{eff}$ & 
$R_\star$ & 
$\log{g}$ & 
$P_\mathrm{rot}$ & 
$P_\mathrm{rot}$ & 
$\alpha_\mathrm{FFD}$ &
$\beta_\mathrm{FFD}$ &
preb. &
ozone depl. &
ozone depl. \\
    & 
    & 
    & 
    & 
max. & 
... &
mean & 
mag & 
 & 
 & 
 &
TESS & 
KELT & 
&
&
chem. &
cons. &
perm. \\
 &                      
 &                        
 &                     
 & 
 ($\mathrm{flux}-1$) & 
 ... &
 (g)                     &          
 & 
 (K)                & 
 ($R_\odot$) &           
 ($\log{ \frac{cm}{s^2} }$)& 
 (d)                    & 
 (d)                     &  
 &
 &
 &
 &
} 
\startdata
\multicolumn{18}{c}{...} \\
167602025 & 2          & 5            & 7          & 0.061     & ... & 2.18E+21     & 11.277  & 5341 & 0.944  & -         & 2.477         & 0.719     & -0.677 & 22.236    & no         & yes               & no                \\
167695269 & 1          & 3            & 3          & 0.512     & ... & 2.74E+20     & 13.001  & 3381 & 0.374  & 4.851     & -             & -         & -      & -         & no         & no                & no                \\
167814740 & 2          & 8            & 11         & 0.589     & ... & 6.62E+20     & 13.243   & 3283 & 0.524  & 4.717     & 0.403         & -         & -0.586 & 18.594    & no         & no                & no                \\
\multicolumn{18}{c}{...} \\
102032397 & 1          & 11           & 13         & 0.093     & ... & 1.68E+21     & 10.593  & 5435 & 1.105  & 4.329 .     & 3.715        & 1.343     & -1.045     & 35.482     & no        & yes                & yes              
    \\    
\multicolumn{18}{c}{...} \\
\hline
\multicolumn{18}{c}{} \\
\multicolumn{18}{c}{
\begin{minipage}{8.7in} 
With one row for each star, the maximum/mean values of the flaring and the stellar values are shown as follows:
the \textit{TIC ID}; 
the number of TESS sectors ($N_\mathrm{sec.}$);
the number of identified outbursts ($N_\mathrm{out.}$);
the number of identified flares ($N_\mathrm{fla.}$);
the maximum and mean flare amplitude (\textit{Amp. max.} and \textit{Amp. mean});
the maximum and mean flare full-width at half-maximum (\textit{FWHM max.} and \textit{FWHM mean});
the maximum bolometric energy of the star's flares ($E_\mathrm{bol.}$ max. and $E_\mathrm{bol.}$ mean);
the maximum and mean mass of coronal mass ejections (\textit{$M_\mathrm{CME}$} max. and \textit{$M_\mathrm{CME}$} mean);
the TESS magnitude (\textit{TESS mag});
the stellar effective temperature ($T_\mathrm{eff}$);
the stellar radius ($R_\star$);
the logarithm of the surface gravity ($\log{g}$);
the rotation period extracted from TESS and KELT data (\textit{TESS rot.} and \textit{KELT rot.}; see Section~\ref{ss:Identifying stellar variability and rotation periods}).
an indication of which star flares enough to potentially trigger prebiotic chemistry on its exoplanets (\textit{preb. chem.}; see Section~\ref{ss:Which flares deliver enough energy for prebiotic chemistry?});
and an indication of whether a star's flaring could lead to ozone depletion on its exoplanets for a conservative (\textit{ozone depl. cons.}) or permissive (\textit{ozone depl. perm.}) threshold (see Section~\ref{ss:Which flares deliver enough energy for prebiotic chemistry?}).
For readability, several columns are hidden in this view.
The machine-readable version is available in the online journal, and contains all described columns. 
It also includes all lower and upper limits values where applicable.
\end{minipage}}\\
\multicolumn{18}{c}{} \\
\enddata
\end{deluxetable*}
\end{longrotatetable}

\bibliography{references}

\begin{thebibliography}{}
\expandafter\ifx\csname natexlab\endcsname\relax\def\natexlab#1{#1}\fi

\bibitem[{Aarnio {et~al.}(2012)Aarnio, Matt, \& Stassun}]{Aarnio2012}
Aarnio, A.~N., Matt, S.~P., \& Stassun, K.~G. 2012, ApJ, 760, 9

\bibitem[{Aarnio {et~al.}(2011)Aarnio, Stassun, Hughes, \&
  McGregor}]{Aarnio2011}
Aarnio, A.~N., Stassun, K.~G., Hughes, W.~J., \& McGregor, S.~L. 2011, Solar
  Physics, 268, 195

\bibitem[{Airapetian {et~al.}(2017)Airapetian, Jackman, Mlynczak, Danchi, \&
  Hunt}]{Airapetian2017}
Airapetian, V.~S., Jackman, C.~H., Mlynczak, M., Danchi, W., \& Hunt, L. 2017,
  Scientific Reports, 7, 14141

\bibitem[{Airapetian \& {others}(2016)}]{Airapetian2016}
Airapetian, V.~S., \& {others}. 2016, Nature Geosci., 9, 452

\bibitem[{Anglada-Escud{\'{e}} {et~al.}(2016)Anglada-Escud{\'{e}}, Amado,
  Barnes, Berdi{\~{n}}as, Butler, Coleman, de~La~Cueva, Dreizler, Endl,
  Giesers, Jeffers, Jenkins, Jones, Kiraga, K{\"{u}}rster,
  L{\'{o}}pez-Gonz{\'{a}}lez, Marvin, Morales, Morin, Nelson, Ortiz, Ofir,
  Paardekooper, Reiners, Rodr{\textbackslash}'{\textbackslash}iguez,
  Rodr{\textbackslash}'{\textbackslash}iguez-L{\'{o}}pez, Sarmiento, Strachan,
  Tsapras, Tuomi, \& Zechmeister}]{Anglada2016}
Anglada-Escud{\'{e}}, G., Amado, P.~J., Barnes, J., {et~al.} 2016, Nature, 536,
  437

\bibitem[{Astudillo-Defru {et~al.}(2017)Astudillo-Defru, Delfosse, Bonfils,
  Forveille, Lovis, \& Rameau}]{Astudillo-Defru2017}
Astudillo-Defru, N., Delfosse, X., Bonfils, X., {et~al.} 2017, A{\&}A, 600, A13

\bibitem[{Atri(2017)}]{Atri2017}
Atri, D. 2017, MNRAS, 465, L34

\bibitem[{Balona {et~al.}(2015)Balona, Broomhall, Kosovichev, Nakariakov, Pugh,
  \& Van~Doorsselaere}]{Balona2015}
Balona, L.~A., Broomhall, A.-M., Kosovichev, A., {et~al.} 2015, MNRAS, 450, 956

\bibitem[{Barnes {et~al.}(2015)Barnes, Meadows, \& Evans}]{Barnes2015}
Barnes, R., Meadows, V.~S., \& Evans, N. 2015, ApJ, 814, 91

\bibitem[{Benz \& G{\"{u}}del(2010)}]{Benz2010}
Benz, A.~O., \& G{\"{u}}del, M. 2010, ARA{\&}A, 48, 241

\bibitem[{Berta {et~al.}(2012)Berta, Irwin, Charbonneau, Burke, \&
  Falco}]{Berta2012}
Berta, Z.~K., Irwin, J., Charbonneau, D., Burke, C.~J., \& Falco, E.~E. 2012,
  AJ, 144, 145

\bibitem[{Berthier {et~al.}(2016)Berthier, Carry, Vachier, Eggl, \&
  Santerne}]{Berthier2016}
Berthier, J., Carry, B., Vachier, F., Eggl, S., \& Santerne, A. 2016, MNRAS,
  458, 3394

\bibitem[{Berthier {et~al.}(2006)Berthier, Vachier, Thuillot, Fernique,
  Ochsenbein, Genova, Lainey, \& Arlot}]{Berthier2006}
Berthier, J., Vachier, F., Thuillot, W., {et~al.} 2006, in Astronomical Society
  of the Pacific Conference Series, Vol. 351, Astronomical Data Analysis
  Software and Systems XV, ed. C.~Gabriel, C.~Arviset, D.~Ponz, \& S.~Enrique,
  367

\bibitem[{Bj{\"{o}}rn \& {others}(2015)}]{Bjoern2015}
Bj{\"{o}}rn, L.~O., \& {others}. 2015, in Photobiology: The Science of Light
  and Life, ed. L.~O. Bj{\"{o}}rn (Springer New York), 415--420

\bibitem[{Borucki {et~al.}(2010)Borucki, Koch, Basri, Batalha, Brown, Caldwell,
  Caldwell, Christensen-Dalsgaard, Cochran, DeVore, Dunham, Dupree, Gautier,
  Geary, Gilliland, Gould, Howell, Jenkins, Kondo, Latham, Marcy, Meibom,
  Kjeldsen, Lissauer, Monet, Morrison, Sasselov, Tarter, Boss, Brownlee, Owen,
  Buzasi, Charbonneau, Doyle, Fortney, Ford, Holman, Seager, Steffen, Welsh,
  Rowe, Anderson, Buchhave, Ciardi, Walkowicz, Sherry, Horch, Isaacson,
  Everett, Fischer, Torres, Johnson, Endl, MacQueen, Bryson, Dotson, Haas,
  Kolodziejczak, Van~Cleve, Chandrasekaran, Twicken, Quintana, Clarke, Allen,
  Li, Wu, Tenenbaum, Verner, Bruhweiler, Barnes, \& Prsa}]{Borucki2010}
Borucki, W.~J., Koch, D., Basri, G., {et~al.} 2010, Science, 327, 977

\bibitem[{Browning(2008)}]{Browning2008}
Browning, M.~K. 2008, ApJ, 676, 1262

\bibitem[{Bryce {et~al.}(2015)Bryce, Horneck, Rabbow, Edwards, \&
  Cockell}]{Bryce2015}
Bryce, C.~C., Horneck, G., Rabbow, E., Edwards, H. G.~M., \& Cockell, C.~S.
  2015, IJAsB, 14, 115

\bibitem[{Carrington(1859)}]{Carrington1859}
Carrington, R.~C. 1859, MNRAS, 20, 13

\bibitem[{Chang {et~al.}(2018)Chang, Lin, Ip, Huang, Hou, Yu, Song, \&
  Luo}]{Chang2018}
Chang, H.~Y., Lin, C.~L., Ip, W.~H., {et~al.} 2018, ApJ, 867, 78

\bibitem[{Cohen {et~al.}(2015)Cohen, Ma, Drake, Glocer, Garraffo, Bell, \&
  Gombosi}]{Cohen2015}
Cohen, O., Ma, Y., Drake, J.~J., {et~al.} 2015, ApJ, 806, 41

\bibitem[{Covey {et~al.}(2008)Covey, Hawley, Bochanski, West, Reid, Golimowski,
  Davenport, Henry, Uomoto, \& Holtzman}]{Covey2008}
Covey, K.~R., Hawley, S.~L., Bochanski, J.~J., {et~al.} 2008, AJ, 136, 1778

\bibitem[{Cullum {et~al.}(2014)Cullum, Stevens, \& Joshi}]{Cullum2014}
Cullum, J., Stevens, D., \& Joshi, M. 2014, Astrobiology, 14, 645

\bibitem[{Cullum \& Stevens(2016)}]{Cullum2016}
Cullum, J., \& Stevens, D.~P. 2016, Proceedings of the National Academy of
  Science, 113, 4278

\bibitem[{Davenport(2016)}]{Davenport2016}
Davenport, J.~R.~A. 2016, ApJ, 829, 23

\bibitem[{Davenport {et~al.}(2014)Davenport, Hawley, Hebb, Wisniewski,
  Kowalski, Johnson, Malatesta, Peraza, Keil, Silverberg, Jansen, Scheffler,
  Berdis, Larsen, \& Hilton}]{Davenport2014}
Davenport, J.~R.~A., Hawley, S.~L., Hebb, L., {et~al.} 2014, ApJ, 797, 122

\bibitem[{Diaz \& Schulze-Makuch(2006)}]{Diaz2006}
Diaz, B., \& Schulze-Makuch, D. 2006, Astrobiology, 6, 332

\bibitem[{Dittmann {et~al.}(2017)Dittmann, Irwin, Charbonneau, Bonfils,
  Astudillo-Defru, Haywood, Berta-Thompson, Newton, Rodriguez, Winters, Tan,
  Almenara, Bouchy, Delfosse, Forveille, Lovis, Murgas, Pepe, Santos, Udry,
  W{\"{u}}nsche, Esquerdo, Latham, \& Dressing}]{Dittmann2017}
Dittmann, J.~A., Irwin, J.~M., Charbonneau, D., {et~al.} 2017, Nature, 544, 333

\bibitem[{Dole(1964)}]{Dole1964}
Dole, S.~H. 1964, {Habitable planets for man} (Blaisdell Publishing Company)

\bibitem[{Doyle {et~al.}(2018)Doyle, Shetye, Antonova, Kolotkov, Srivastava,
  Stangalini, Gupta, Avramova, \& Mathioudakis}]{Doyle2018}
Doyle, J.~G., Shetye, J., Antonova, A.~E., {et~al.} 2018, MNRAS, 475,
  doi:10.1093/mnras/sty032

\bibitem[{Durney(1993)}]{Durney93}
Durney, B.~R. 1993, ApJ, 407, 367

\bibitem[{Estrela \& Valio(2018)}]{Estrela2018}
Estrela, R., \& Valio, A. 2018, Astrobiology, 18, 1414

\bibitem[{Foreman-Mackey(2016)}]{Foreman-Mackey2016}
Foreman-Mackey, D. 2016, JOSS, 24, doi:10.21105/joss.00024

\bibitem[{Foreman-Mackey {et~al.}(2017)Foreman-Mackey, Agol, Ambikasaran, \&
  Angus}]{Foreman-Mackey2017}
Foreman-Mackey, D., Agol, E., Ambikasaran, S., \& Angus, R. 2017, {celerite:
  Scalable 1D Gaussian Processes in C++, Python, and Julia}, Astrophysics
  Source Code Library, ,

\bibitem[{Foreman-Mackey {et~al.}(2013)Foreman-Mackey, Hogg, Lang, \&
  Goodman}]{Foreman-Mackey2013}
Foreman-Mackey, D., Hogg, D.~W., Lang, D., \& Goodman, J. 2013, PASP, 125, 306

\bibitem[{France {et~al.}(2016)France, Parke~Loyd, Youngblood, Brown,
  Schneider, Hawley, Froning, Linsky, Roberge, Buccino, Davenport, Fontenla,
  Kaltenegger, Kowalski, Mauas, Miguel, Redfield, Rugheimer, Tian, Vieytes,
  Walkowicz, \& Weisenburger}]{France2016}
France, K., Parke~Loyd, R.~O., Youngblood, A., {et~al.} 2016, ApJ, 820, 89

\bibitem[{Gershberg(1972)}]{Gershberg1972}
Gershberg, R.~E. 1972, APSS, 19, 75

\bibitem[{Gillon {et~al.}(2017)Gillon, Triaud, Demory, Jehin, Agol, Deck,
  Lederer, de~Wit, Burdanov, Ingalls, Bolmont, Leconte, Raymond, Selsis,
  Turbet, Barkaoui, Burgasser, Burleigh, Carey, Chaushev, Copperwheat, Delrez,
  Fernandes, Holdsworth, Kotze, Van~Grootel, Almleaky, Benkhaldoun, Magain, \&
  Queloz}]{Gillon2017}
Gillon, M., Triaud, A.~H.~M.~J., Demory, B.-O., {et~al.} 2017, Nature, 542, 456

\bibitem[{Grenfell {et~al.}(2012)Grenfell, Grie{\ss}meier, von Paris, Patzer,
  Lammer, Stracke, Gebauer, Schreier, \& Rauer}]{Grenfell2012}
Grenfell, J.~L., Grie{\ss}meier, J.-M., von Paris, P., {et~al.} 2012,
  Astrobiology, 12, 1109

\bibitem[{G{\"{u}}nther \& Daylan(2019)}]{allesfitter-code}
G{\"{u}}nther, M.~N., \& Daylan, T. 2019, {Allesfitter: Flexible Star and
  Exoplanet Inference From Photometry and Radial Velocity}, Astrophysics Source
  Code Library, ,

\bibitem[{G{\"{u}}nther \& Daylan(2020)}]{allesfitter-paper}
---. 2020, arXiv e-prints, arXiv:2003.14371

\bibitem[{Hawley {et~al.}(2014)Hawley, Davenport, Kowalski, Wisniewski, Hebb,
  Deitrick, \& Hilton}]{Hawley2014}
Hawley, S.~L., Davenport, J.~R.~A., Kowalski, A.~F., {et~al.} 2014, ApJ, 797,
  121

\bibitem[{Hawley \& Pettersen(1991)}]{Hawley1991}
Hawley, S.~L., \& Pettersen, B.~R. 1991, ApJ, 378, 725

\bibitem[{Henry {et~al.}(1994)Henry, Kirkpatrick, \& Simons}]{Henry1994}
Henry, T.~J., Kirkpatrick, J.~D., \& Simons, D.~A. 1994, AJ, 108, 1437

\bibitem[{Hodgson(1859)}]{Hodgson1859}
Hodgson, R. 1859, MNRAS, 20, 15

\bibitem[{Howard {et~al.}(2018)Howard, Tilley, Corbett, Youngblood, Loyd,
  Ratzloff, Law, Fors, del Ser, Shkolnik, Ziegler, Goeke, Pietraallo, \&
  Haislip}]{Howard2018}
Howard, W.~S., Tilley, M.~A., Corbett, H., {et~al.} 2018, ApJ, 860, L30

\bibitem[{Huang(1959)}]{Huang1959}
Huang, S.-S. 1959, AmSci, 47, 397

\bibitem[{Hunter(2007)}]{Hunter2007}
Hunter, J.~D. 2007, CSE, 9, 90

\bibitem[{Jackman {et~al.}(2019)Jackman, Wheatley, Pugh, Kolotkov, Broomhall,
  Kennedy, Murphy, Raddi, Burleigh, Casewell, Eigm{\"{u}}ller, Gillen,
  G{\"{u}}nther, Jenkins, Louden, McCormac, Raynard, Poppenhaeger, Udry,
  Watson, \& West}]{Jackman2019}
Jackman, J.~A., Wheatley, P.~J., Pugh, C.~E., {et~al.} 2019, MNRAS, 482, 5553

\bibitem[{Jackman {et~al.}(2018)Jackman, Wheatley, Pugh, G{\"{a}}nsicke,
  Gillen, Broomhall, Armstrong, Burleigh, Chaushev, Eigm{\"{u}}ller, Erikson,
  Goad, Grange, G{\"{u}}nther, Jenkins, McCormac, Raynard, Thompson, Udry,
  Walker, Watson, \& West}]{Jackman2018}
Jackman, J. A.~G., Wheatley, P.~J., Pugh, C.~E., {et~al.} 2018, MNRAS, 477,
  4655

\bibitem[{Jenkins(2002)}]{Jenkins2002}
Jenkins, J.~M. 2002, ApJ, 575, 493

\bibitem[{Jenkins(2017)}]{Jenkins2017}
---. 2017, {Kepler Data Processing Handbook}, Tech. rep.

\bibitem[{Jenkins {et~al.}(2010)Jenkins, Chandrasekaran, McCauliff, Caldwell,
  Tenenbaum, Li, Klaus, Cote, \& Middour}]{Jenkins2010}
Jenkins, J.~M., Chandrasekaran, H., McCauliff, S.~D., {et~al.} 2010, in
  {\textbackslash}procspie, Vol. 7740, Software and Cyberinfrastructure for
  Astronomy, 77400D

\bibitem[{Jenkins {et~al.}(2016)Jenkins, Twicken, McCauliff, Campbell,
  Sanderfer, Lung, Mansouri-Samani, Girouard, Tenenbaum, Klaus, Smith,
  Caldwell, Chacon, Henze, Heiges, Latham, Morgan, Swade, Rinehart, \&
  Vanderspek}]{Jenkins2016}
Jenkins, J.~M., Twicken, J.~D., McCauliff, S., {et~al.} 2016, in
  {\textbackslash}procspie, Vol. 9913, Software and Cyberinfrastructure for
  Astronomy IV, 99133E

\bibitem[{Jones {et~al.}(2001)Jones, Oliphant, Peterson, \&
  {Others}}]{Jones2001}
Jones, E., Oliphant, T., Peterson, P., \& {Others}. 2001, {SciPy: Open Source
  Scientific Tools for Python, 2001 (http://www.scipy.org/)}, ,

\bibitem[{Kaltenegger \& Traub(2009)}]{Kaltenegger2009}
Kaltenegger, L., \& Traub, W.~A. 2009, ApJ, 698, 519

\bibitem[{Kane {et~al.}(2016)Kane, Hill, Kasting, Kopparapu, Quintana, Barclay,
  Batalha, Borucki, Ciardi, Haghighipour, Hinkel, Kaltenegger, Selsis, \&
  Torres}]{Kane2016}
Kane, S.~R., Hill, M.~L., Kasting, J.~F., {et~al.} 2016, ApJ, 830, 1

\bibitem[{Kass \& Raftery(1995)}]{Kass1995}
Kass, R.~E., \& Raftery, A.~E. 1995, J. Am. Stat. Assoc., 90, 773

\bibitem[{Kasting {et~al.}(2014)Kasting, Kopparapu, Ramirez, \&
  Harman}]{Kasting2014}
Kasting, J.~F., Kopparapu, R., Ramirez, R.~M., \& Harman, C.~E. 2014,
  Proceedings of the National Academy of Science, 111, 12641

\bibitem[{Kasting {et~al.}(1993)Kasting, Whitmire, \& Reynolds}]{Kasting1993}
Kasting, J.~F., Whitmire, D.~P., \& Reynolds, R.~T. 1993, Icarus, 101, 108

\bibitem[{Kay {et~al.}(2016)Kay, Opher, \& Kornbleuth}]{Kay2016}
Kay, C., Opher, M., \& Kornbleuth, M. 2016, ApJ, 826, 195

\bibitem[{Kiang {et~al.}(2007)Kiang, Segura, Tinetti, {Govindjee}, Blankenship,
  Cohen, Siefert, Crisp, \& Meadows}]{Kiang2007}
Kiang, N.~Y., Segura, A., Tinetti, G., {et~al.} 2007, Astrobiology, 7, 252

\bibitem[{Kipping {et~al.}(2017)Kipping, Cameron, Hartman, Davenport, Matthews,
  Sasselov, Rowe, Siverd, Chen, Sandford, Bakos, Jord{\'{a}}n, Bayliss,
  Henning, Mancini, Penev, Csubry, Bhatti, Da~Silva~Bento, Guenther, Kuschnig,
  Moffat, Rucinski, \& Weiss}]{Kipping2017}
Kipping, D.~M., Cameron, C., Hartman, J.~D., {et~al.} 2017, AJ, 153, 93

\bibitem[{Kopparapu {et~al.}(2014)Kopparapu, Ramirez, SchottelKotte, Kasting,
  Domagal-Goldman, \& Eymet}]{Kopparapu2014}
Kopparapu, R.~K., Ramirez, R.~M., SchottelKotte, J., {et~al.} 2014, ApJ, 787,
  L29

\bibitem[{Kopparapu {et~al.}(2013)Kopparapu, Ramirez, Kasting, Eymet, Robinson,
  Mahadevan, Terrien, Domagal-Goldman, Meadows, \& Deshpande}]{Kopparapu2013}
Kopparapu, R.~K., Ramirez, R., Kasting, J.~F., {et~al.} 2013, ApJ, 765, 131

\bibitem[{Kowalski {et~al.}(2009)Kowalski, Hawley, Hilton, Becker, West,
  Bochanski, \& Sesar}]{Kowalski2009}
Kowalski, A.~F., Hawley, S.~L., Hilton, E.~J., {et~al.} 2009, AJ, 138, 633

\bibitem[{Kowalski {et~al.}(2013)Kowalski, Hawley, Wisniewski, Osten, Hilton,
  Holtzman, Schmidt, \& Davenport}]{Kowalski2013}
Kowalski, A.~F., Hawley, S.~L., Wisniewski, J.~P., {et~al.} 2013, ApJS, 207, 15

\bibitem[{Lacy {et~al.}(1976)Lacy, Moffett, \& Evans}]{Lacy1976}
Lacy, C.~H., Moffett, T.~J., \& Evans, D.~S. 1976, APJS, 30, 85

\bibitem[{Lammer {et~al.}(2007)Lammer, Lichtenegger, Kulikov, Grie{\ss}meier,
  Terada, Erkaev, Biernat, Khodachenko, Ribas, Penz, \& Selsis}]{Lammer2007}
Lammer, H., Lichtenegger, H.~I.~M., Kulikov, Y.~N., {et~al.} 2007,
  Astrobiology, 7, 185

\bibitem[{Law {et~al.}(2015)Law, Fors, Ratzloff, Wulfken, Kavanaugh, Sitar,
  Pruett, Birchard, Barlow, Cannon, Cenko, Dunlap, Kraus, \&
  Maccarone}]{Law2015}
Law, N.~M., Fors, O., Ratzloff, J., {et~al.} 2015, PASP, 127, 234

\bibitem[{Lingam \& Loeb(2017)}]{Lingam2017superflares}
Lingam, M., \& Loeb, A. 2017, ApJ, 848, 41

\bibitem[{Lomb(1976)}]{Lomb:1976}
Lomb, N.~R. 1976, Ap{\&}SS, 39, 447

\bibitem[{Maehara {et~al.}(2015)Maehara, Shibayama, Notsu, Notsu, Honda,
  Nogami, \& Shibata}]{Maehara2015}
Maehara, H., Shibayama, T., Notsu, Y., {et~al.} 2015, Earth, Planets and Space,
  67, 59

\bibitem[{Maehara {et~al.}(2012)Maehara, Shibayama, Notsu, Notsu, Nagao,
  Kusaba, Honda, Nogami, \& Shibata}]{Maehara2012}
Maehara, H., Shibayama, T., Notsu, S., {et~al.} 2012, Nature, 485, 478

\bibitem[{Maxted(2016)}]{Maxted2016}
Maxted, P.~F.~L. 2016, A{\&}A, 591, A111

\bibitem[{Meadows \& Barnes(2018)}]{Meadows2018}
Meadows, V.~S., \& Barnes, R.~K. 2018, in Handbook of Exoplanets, ISBN
  978-3-319-55332-0. Springer International Publishing AG, part of Springer
  Nature, 2018, id.57, 57

\bibitem[{Moffatt(1978)}]{Moffatt1978}
Moffatt, H.~K. 1978, {Magnetic field generation in electrically conducting
  fluids} (Cambridge University Press)

\bibitem[{Mondrik {et~al.}(2019)Mondrik, Newton, Charbonneau, \&
  Irwin}]{Mondrik2019}
Mondrik, N., Newton, E., Charbonneau, D., \& Irwin, J. 2019, ApJ, 870, 10

\bibitem[{Mullan \& Bais(2018)}]{Mullan2018}
Mullan, D.~J., \& Bais, H.~P. 2018, ApJ, 865, 101

\bibitem[{Newton {et~al.}(2017)Newton, Irwin, Charbonneau, Berlind, Calkins, \&
  Mink}]{Newton2017}
Newton, E.~R., Irwin, J., Charbonneau, D., {et~al.} 2017, ApJ, 834, 85

\bibitem[{Newton {et~al.}(2016)Newton, Irwin, Charbonneau, Berta-Thompson, \&
  Dittmann}]{Newton2016a}
Newton, E.~R., Irwin, J., Charbonneau, D., Berta-Thompson, Z.~K., \& Dittmann,
  J.~A. 2016, ApJ, 821, L19

\bibitem[{Newton {et~al.}(2018)Newton, Mondrik, Irwin, Winters, \&
  Charbonneau}]{Newton2018}
Newton, E.~R., Mondrik, N., Irwin, J., Winters, J.~G., \& Charbonneau, D. 2018,
  AJ, 156, 217

\bibitem[{Noyes {et~al.}(1984)Noyes, Hartmann, Baliunas, Duncan, \&
  Vaughan}]{Noyes1984}
Noyes, R.~W., Hartmann, L.~W., Baliunas, S.~L., Duncan, D.~K., \& Vaughan,
  A.~H. 1984, ApJ, 279, 763

\bibitem[{Nutzman \& Charbonneau(2008)}]{Nutzman2008}
Nutzman, P., \& Charbonneau, D. 2008, PASP, 120, 317

\bibitem[{Oelkers {et~al.}(2018)Oelkers, Rodriguez, Stassun, Pepper, Somers,
  Kafka, Stevens, Beatty, Siverd, Lund, Kuhn, James, \& Gaudi}]{Oelkers2018}
Oelkers, R.~J., Rodriguez, J.~E., Stassun, K.~G., {et~al.} 2018, AJ, 155, 39

\bibitem[{O'Malley-James \& Kaltenegger(2018)}]{OmalleyJames2018}
O'Malley-James, J.~T., \& Kaltenegger, L. 2018, MNRAS, 481, 2487

\bibitem[{Osten \& Wolk(2015)}]{Osten2015}
Osten, R.~A., \& Wolk, S.~J. 2015, ApJ, 809, 79

\bibitem[{P{\'{a}}l {et~al.}(2018)P{\'{a}}l, Moln{\'{a}}r, \& Kiss}]{Pal2018}
P{\'{a}}l, A., Moln{\'{a}}r, L., \& Kiss, C. 2018, PASP, 130, 114503

\bibitem[{Parker(1979)}]{Parker1979}
Parker, E.~N. 1979, {Cosmical magnetic fields. Their origin and their activity}

\bibitem[{Parnell \& Jupp(2000)}]{Parnell2000}
Parnell, C.~E., \& Jupp, P.~E. 2000, ApJ, 529, 554

\bibitem[{Pecaut \& Mamajek(2013)}]{Pecaut2013}
Pecaut, M.~J., \& Mamajek, E.~E. 2013, ApJS, 208, 9

\bibitem[{Pierrehumbert \& Gaidos(2011)}]{Pierrehumbert2011}
Pierrehumbert, R., \& Gaidos, E. 2011, ApJ, 734, L13

\bibitem[{Ramirez \& Kaltenegger(2014)}]{Ramirez2014}
Ramirez, R.~M., \& Kaltenegger, L. 2014, ApJ, 797, L25

\bibitem[{Ranjan {et~al.}(2017)Ranjan, Wordsworth, \& Sasselov}]{Ranjan2017}
Ranjan, S., Wordsworth, R., \& Sasselov, D.~D. 2017, ApJ, 843, 110

\bibitem[{Reid {et~al.}(2004)Reid, Cruz, Allen, Mungall, Kilkenny, Liebert,
  Hawley, Fraser, Covey, Lowrance, Kirkpatrick, \& Burgasser}]{Reid2004}
Reid, I.~N., Cruz, K.~L., Allen, P., {et~al.} 2004, AJ, 128, 463

\bibitem[{Ricker {et~al.}(2014)Ricker, Winn, Vanderspek, Latham, Bakos, Bean,
  Berta-Thompson, Brown, Buchhave, Butler, Butler, Chaplin, Charbonneau,
  Christensen-Dalsgaard, Clampin, Deming, Doty, De~Lee, Dressing, Dunham, Endl,
  Fressin, Ge, Henning, Holman, Howard, Ida, Jenkins, Jernigan, Johnson,
  Kaltenegger, Kawai, Kjeldsen, Laughlin, Levine, Lin, Lissauer, MacQueen,
  Marcy, McCullough, Morton, Narita, Paegert, Palle, Pepe, Pepper, Quirrenbach,
  Rinehart, Sasselov, Sato, Seager, Sozzetti, Stassun, Sullivan, Szentgyorgyi,
  Torres, Udry, \& Villasenor}]{Ricker2014}
Ricker, G.~R., Winn, J.~N., Vanderspek, R., {et~al.} 2014, in SPIE Conf.
  Series, Vol. 9143, 20

\bibitem[{Rimmer {et~al.}(2018)Rimmer, Xu, Thompson, Gillen, Sutherland, \&
  Queloz}]{Rimmer2018}
Rimmer, P.~B., Xu, J., Thompson, S.~J., {et~al.} 2018, Science Advances, 4,
  eaar3302

\bibitem[{Rodono {et~al.}(1986)Rodono, Cutispoto, Pazzani, Catalano, Byrne,
  Doyle, Butler, Andrews, Blanco, Marilli, Linsky, Scaltriti, Busso, Cellino,
  Hopkins, Okazaki, Hayashi, Zeilik, Helston, Henson, Smith, \&
  Simon}]{Rodono1986}
Rodono, M., Cutispoto, G., Pazzani, V., {et~al.} 1986, A{\&}A, 165, 135

\bibitem[{Scalo {et~al.}(2007)Scalo, Kaltenegger, Segura, Fridlund, Ribas,
  Kulikov, Grenfell, Rauer, Odert, Leitzinger, Selsis, Khodachenko, Eiroa,
  Kasting, \& Lammer}]{Scalo2007}
Scalo, J., Kaltenegger, L., Segura, A.~G., {et~al.} 2007, Astrobiology, 7, 85

\bibitem[{Scargle(1982)}]{Scargle:1982}
Scargle, J.~D. 1982, ApJ, 263, 835

\bibitem[{Seager {et~al.}(2013)Seager, Bains, \& Hu}]{Seager2013}
Seager, S., Bains, W., \& Hu, R. 2013, ApJ, 777, 95

\bibitem[{Segura {et~al.}(2003)Segura, Krelove, Kasting, Sommerlatt, Meadows,
  Crisp, Cohen, \& Mlawer}]{Segura2003}
Segura, A., Krelove, K., Kasting, J.~F., {et~al.} 2003, Astrobiology, 3, 689

\bibitem[{Segura {et~al.}(2010)Segura, Walkowicz, Meadows, Kasting, \&
  Hawley}]{Segura2010}
Segura, A., Walkowicz, L.~M., Meadows, V., Kasting, J., \& Hawley, S. 2010,
  Astrobiology, 10, 751

\bibitem[{Shibata \& Takasao(2016)}]{Shibata2016}
Shibata, K., \& Takasao, S. 2016, in Astrophysics and Space Science Library,
  Vol. 427, Magnetic Reconnection: Concepts and Applications, ed. W.~Gonzalez
  \& E.~Parker, 373

\bibitem[{Shibayama {et~al.}(2013)Shibayama, Maehara, Notsu, Notsu, Nagao,
  Honda, Ishii, Nogami, \& Shibata}]{Shibayama2013}
Shibayama, T., Maehara, H., Notsu, S., {et~al.} 2013, ApJS, 209, 5

\bibitem[{Shields {et~al.}(2016)Shields, Ballard, \& Johnson}]{Shields2016}
Shields, A.~L., Ballard, S., \& Johnson, J.~A. 2016, PhysRep, 663, 1

\bibitem[{Simon \& Drake(1989)}]{Simon1989}
Simon, T., \& Drake, S.~A. 1989, ApJ, 346, 303

\bibitem[{Smith {et~al.}(2004)Smith, Scalo, \& Wheeler}]{Smith2004}
Smith, D.~S., Scalo, J., \& Wheeler, J.~C. 2004, Icarus, 171, 229

\bibitem[{Smith {et~al.}(2012)Smith, Stumpe, Van~Cleve, Jenkins, Barclay,
  Fanelli, Girouard, Kolodziejczak, McCauliff, Morris, \& Twicken}]{Smith2012}
Smith, J.~C., Stumpe, M.~C., Van~Cleve, J.~E., {et~al.} 2012, PASP, 124, 1000

\bibitem[{Speagle(2020)}]{Speagle2020}
Speagle, J.~S. 2020, MNRAS, doi:10.1093/mnras/staa278

\bibitem[{Stassun {et~al.}(2011)Stassun, Hebb, Covey, West, Irwin, Jackson,
  Jardine, Morin, Mullan, \& Reid}]{Stassun2011}
Stassun, K.~G., Hebb, L., Covey, K., {et~al.} 2011, in Astronomical Society of
  the Pacific Conference Series, Vol. 448, 16{\$}{\^{}}{\{}th{\}}{\$} Cambridge
  Workshop on Cool Stars, Stellar Systems, and the Sun. ASP Conference Series,
  Vol. 448, proceedings of a conference held August 28- September 2, 2010 at
  the University of Washington, Seattle, Washington. Edited by Christopher M.
  Johns-K, ed. C.~Johns-Krull, M.~K. Browning, \& A.~A. West, 505

\bibitem[{Stellingwerf(1978)}]{Stellingwerf1978}
Stellingwerf, R.~F. 1978, ApJ, 224, 953

\bibitem[{Stelzer {et~al.}(2016)Stelzer, Damasso, Scholz, \&
  Matt}]{Stelzer2016}
Stelzer, B., Damasso, M., Scholz, A., \& Matt, S.~P. 2016, MNRAS, 463, 1844

\bibitem[{Stetson(1996)}]{Stetson:1996}
Stetson, P.~B. 1996, PASP, 108, 851

\bibitem[{Stumpe {et~al.}(2014)Stumpe, Smith, Catanzarite, Van~Cleve, Jenkins,
  Twicken, \& Girouard}]{Stumpe2014}
Stumpe, M.~C., Smith, J.~C., Catanzarite, J.~H., {et~al.} 2014, PASP, 126, 100

\bibitem[{Sutherland(2015)}]{Sutherland2015}
Sutherland, J. 2015, in European Planetary Science Congress, 2015--1

\bibitem[{Szab{\'{o}} {et~al.}(2015)Szab{\'{o}}, S{\'{a}}rneczky, Szab{\'{o}},
  P{\'{a}}l, Kiss, Cs{\'{a}}k, Ill{\'{e}}s, R{\'{a}}cz, \& Kiss}]{Szabo2015}
Szab{\'{o}}, R., S{\'{a}}rneczky, K., Szab{\'{o}}, G.~M., {et~al.} 2015, AJ,
  149, 112

\bibitem[{Teachey \& Kipping(2018)}]{Teachey2018}
Teachey, A., \& Kipping, D.~M. 2018, Science Advances, 4, eaav1784

\bibitem[{Tilley {et~al.}(2019)Tilley, Segura, Meadows, Hawley, \&
  Davenport}]{Tilley2019}
Tilley, M.~A., Segura, A., Meadows, V., Hawley, S., \& Davenport, J. 2019,
  Astrobiology, 19, 64

\bibitem[{van~der Walt {et~al.}(2011)van~der Walt, Colbert, \&
  Varoquaux}]{vanderWalt2011}
van~der Walt, S., Colbert, S.~C., \& Varoquaux, G. 2011, CSE, 13, 22

\bibitem[{Van~Doorsselaere {et~al.}(2017)Van~Doorsselaere, Shariati, \&
  Debosscher}]{VanDoorsselaere2017}
Van~Doorsselaere, T., Shariati, H., \& Debosscher, J. 2017, ApJSS, 232, 26

\bibitem[{van Rossum(1995)}]{Rossum1995}
van Rossum, G. 1995, {Python tutorial}, Tech. Rep. CS-R9526, Centrum voor
  Wiskunde en Informatica (CWI), Amsterdam

\bibitem[{Vanderburg {et~al.}(2016)Vanderburg, Plavchan, Johnson, Ciardi,
  Swift, \& Kane}]{Vanderburg2016}
Vanderburg, A., Plavchan, P., Johnson, J.~A., {et~al.} 2016, MNRAS, 459, 3565

\bibitem[{Venot {et~al.}(2016)Venot, Rocchetto, Carl, Roshni~Hashim, \&
  Decin}]{Venot2016}
Venot, O., Rocchetto, M., Carl, S., Roshni~Hashim, A., \& Decin, L. 2016, ApJ,
  830, 77

\bibitem[{Vin{\'{i}}cius {et~al.}(2018)Vin{\'{i}}cius, Barentsen, Hedges,
  Gully-Santiago, \& Cody}]{lightkurve}
Vin{\'{i}}cius, Z., Barentsen, G., Hedges, C., Gully-Santiago, M., \& Cody,
  A.~M. 2018, {KeplerGO/lightkurve}, , , doi:10.5281/zenodo.1181928

\bibitem[{Walkowicz {et~al.}(2011)Walkowicz, Basri, Batalha, Gilliland,
  Jenkins, Borucki, Koch, Caldwell, Dupree, Latham, Meibom, Howell, Brown, \&
  Bryson}]{Walkowicz2011}
Walkowicz, L.~M., Basri, G., Batalha, N., {et~al.} 2011, AJ, 141, 50

\bibitem[{West {et~al.}(2008)West, Hawley, Bochanski, Covey, Reid, Dhital,
  Hilton, \& Masuda}]{West2008}
West, A.~A., Hawley, S.~L., Bochanski, J.~J., {et~al.} 2008, AJ, 135, 785

\bibitem[{West {et~al.}(2015)West, Weisenburger, Irwin, Berta-Thompson,
  Charbonneau, Dittmann, \& Pineda}]{West2015}
West, A.~A., Weisenburger, K.~L., Irwin, J., {et~al.} 2015, ApJ, 812, 3

\bibitem[{Wheatley {et~al.}(2018)Wheatley, West, Goad, Jenkins, Pollacco,
  Queloz, Rauer, Udry, Watson, Chazelas, Eigm{\"{u}}ller, Lambert, Genolet,
  McCormac, Walker, Armstrong, Bayliss, Bento, Bouchy, Burleigh, Cabrera,
  Casewell, Chaushev, Chote, Csizmadia, Erikson, Faedi, Foxell, G{\"{a}}nsicke,
  Gillen, Grange, G{\"{u}}nther, Hodgkin, Jackman, Jord{\'{a}}n, Louden,
  Metrailler, Moyano, Nielsen, Osborn, Poppenhaeger, Raddi, Raynard, Smith,
  Soto, \& Titz-Weider}]{Wheatley2018}
Wheatley, P., West, R., Goad, M., {et~al.} 2018, MNRAS, 475,
  doi:10.1093/mnras/stx2836

\bibitem[{Wright {et~al.}(2018)Wright, Newton, Williams, Drake, \&
  Yadav}]{Wright2018}
Wright, N.~J., Newton, E.~R., Williams, P.~K.~G., Drake, J.~J., \& Yadav, R.~K.
  2018, MNRAS, 479, 2351

\bibitem[{Xu {et~al.}(2018)Xu, Ritson, Ranjan, Todd, Sasselov, \&
  Sutherland}]{Xu2018}
Xu, J., Ritson, D.~J., Ranjan, S., {et~al.} 2018, Chemical Communications, 54,
  5566

\bibitem[{York {et~al.}(2000)York, Adelman, Anderson John~E., Anderson, Annis,
  Bahcall, Bakken, Barkhouser, Bastian, Berman, Boroski, Bracker, Briegel,
  Briggs, Brinkmann, Brunner, Burles, Carey, Carr, Castander, Chen, Colestock,
  Connolly, Crocker, Csabai, Czarapata, Davis, Doi, Dombeck, Eisenstein,
  Ellman, Elms, Evans, Fan, Federwitz, Fiscelli, Friedman, Frieman, Fukugita,
  Gillespie, Gunn, Gurbani, de~Haas, Haldeman, Harris, Hayes, Heckman,
  Hennessy, Hindsley, Holm, Holmgren, Huang, Hull, Husby, Ichikawa, Ichikawa,
  Ivezi{\'{c}}, Kent, Kim, Kinney, Klaene, Kleinman, Kleinman, Knapp, Korienek,
  Kron, Kunszt, Lamb, Lee, Leger, Limmongkol, Lindenmeyer, Long, Loomis,
  Loveday, Lucinio, Lupton, MacKinnon, Mannery, Mantsch, Margon, McGehee,
  McKay, Meiksin, Merelli, Monet, Munn, Narayanan, Nash, Neilsen, Neswold,
  Newberg, Nichol, Nicinski, Nonino, Okada, Okamura, Ostriker, Owen, Pauls,
  Peoples, Peterson, Petravick, Pier, Pope, Pordes, Prosapio, Rechenmacher,
  Quinn, Richards, Richmond, Rivetta, Rockosi, Ruthmansdorfer, Sand~ford,
  Schlegel, Schneider, Sekiguchi, Sergey, Shimasaku, Siegmund, Smee, Smith,
  Snedden, Stone, Stoughton, Strauss, Stubbs, SubbaRao, Szalay, Szapudi,
  Szokoly, Thakar, Tremonti, Tucker, Uomoto, Vanden~Berk, Vogeley, Waddell,
  Wang, Watanabe, Weinberg, Yanny, Yasuda, \& {SDSS Collaboration}}]{York2000}
York, D.~G., Adelman, J., Anderson John~E., J., {et~al.} 2000, AJ, 120, 1579

\bibitem[{Youngblood {et~al.}(2017)Youngblood, France, Loyd, Brown, Mason,
  Schneider, Tilley, Berta-Thompson, Buccino, Froning, Hawley, Linsky, Mauas,
  Redfield, Kowalski, Miguel, Newton, Rugheimer, Segura, Roberge, \&
  Vieytes}]{Youngblood2017}
Youngblood, A., France, K., Loyd, R.~O.~P., {et~al.} 2017, ApJ, 843, 31

\bibitem[{Zhan {et~al.}(2019)Zhan, G{\"{u}}nther, Rappaport, Ol{\'{a}}h, Mann,
  Levine, Winn, Dai, Zhou, Huang, Bouma, Ireland, Ricker, Vanderspek, Latham,
  Seager, Jenkins, Caldwell, Doty, Essack, Furesz, Leidos, Rowden, Smith,
  Stassun, \& Vezie}]{Zhan2019}
Zhan, Z., G{\"{u}}nther, M., Rappaport, S., {et~al.} 2019, ApJ, 876,
  doi:10.3847/1538-4357/ab158c

\end{thebibliography}


\end{document}